On the abundance of non-cometary HCN on Jupiter

Julianne I. Moses[†]

Space Science Institute

1602 Old Orchard Ln

Seabrook, TX 77586, USA

and

Channon Visscher

Lunar and Planetary Institute

3600 Bay Area Blvd.

Houston, TX 77058, USA

and

Thomas C. Keane

Dept. of Chemistry and Biochemistry

Russell Sage College

Troy, NY 12180, USA

and

Aubrey Sperier

Dept. of Chemistry and Physics

University of St. Thomas

Houston, TX 77006-4626, USA

[†]Author to whom correspondence should be addressed.

Proposed running head:

On the abundance of non-cometary HCN on Jupiter

Send correspondence to:

Julianne I. Moses
Space Science Institute
1602 Old Orchard Ln
Seabrook, TX 77586
USA

jmoses@spacescience.org
281.474.9996 (phone)

## Abstract

Using one-dimensional thermochemical/photochemical kinetics and transport models, we examine the chemistry of nitrogen-bearing species in the Jovian troposphere in an attempt to explain the low observational upper limit for HCN. We track the dominant mechanisms for interconversion of $N_2$-$NH_3$ and HCN-$NH_3$ in the deep, high-temperature troposphere and predict the rate-limiting step for the quenching of HCN at cooler tropospheric altitudes. Consistent with other investigations that were based solely on time-scale arguments, our models suggest that transport-induced quenching of thermochemically derived HCN leads to very small predicted mole fractions of hydrogen cyanide in Jupiter's upper troposphere. By the same token, photochemical production of HCN is ineffective in Jupiter's troposphere: $CH_4$-$NH_3$ coupling is inhibited by the physical separation of the $CH_4$ photolysis region in the upper stratosphere from the $NH_3$ photolysis and condensation region in the troposphere, and $C_2H_2$-$NH_3$ coupling is inhibited by the low tropospheric abundance of $C_2H_2$. The upper limits from infrared and submillimeter observations can be used to place constraints on the production of HCN and other species from lightning and thundershock sources.

## 1. Introduction

Although hydrogen cyanide (HCN) was detected in the Jovian stratosphere following the Comet Shoemaker-Levy 9 impacts,[1-9] no convincing observational evidence exists for the presence of non-cometary HCN in Jupiter's troposphere. Tentative detections of HCN from the 1960's and 1970's have all been discounted.[10] The most credible report of the detection of non-cometary HCN on Jupiter resulted from ground-based 13.5 $\mu$m observations of three spectral lines by Tokunaga *et al.*;[11] however, this detection is now considered dubious due to the lack of confirmation from subsequent infrared, sub-millimeter, and millimeter observations.[12-15] Bézard *et al.*[13] suggest that one of the purported HCN absorption lines identified by Tokunaga *et al.*[11] is actually an expected "valley" between two nearby $C_2H_2$ emission lines, a second absorption line is much narrower than it would be if caused by HCN (and is likely an instrument artifact), and the identification of the third line is suspect due to uncertainties in the position (frequency) of the feature. From recent 850-$\mu$m observations, a strict upper limit of 0.93 ppb has been placed on the Jovian tropospheric HCN mole fraction, assuming HCN condenses in the upper troposphere, or as small as 0.16 ppb if the HCN is assumed to be uniformly mixed throughout the troposphere and stratosphere.[15]

The production of hydrogen cyanide and other nitrogen-bearing organics in reducing atmospheres such as that of Jupiter has attracted considerable interest in the past half century due to prebiological chemistry implications[16] and to the long-standing puzzle of the cloud coloring agents on Jupiter, for which it has been suggested that HCN polymers could play a role.[17,18] The HCN abundance is expected to be negligible in thermochemical equilibrium at the cold atmospheric levels that can be probed by remote-sensing observations,[19] but several disequilibrium processes could supply HCN to the Jovian troposphere. These processes include rapid transport from the deep troposphere,[10,19-21] photochemical processing of $CH_3NH_2$ dredged up from the deep atmosphere,[10] lightning and related processes in thunderstorms,[17,22-26] coupled $CH_4$-$NH_3$ photochemistry,[27-34] and coupled $C_2H_2$-$NH_3$ photochemistry.[35-40]

Kaye and Strobel[35] and Lewis and Fegley[10] have evaluated the various disequilibrium processes for HCN production and conclude that $NH_3$-$C_2H_2$ photochemical coupling is the most plausible mechanism for producing HCN on Jupiter — HCN production from the chemistry of hot H atoms released from $NH_3$ (or $H_2S$) photolysis in the presence of methane is inhibited under Jovian conditions because of rapid hot-atom thermalization from collisions with $H_2$,[31,41,42] photochemical production of the HCN precursor $CH_3NH_2$ is inhibited due to an insufficient source of $CH_3$ in the $NH_3$ photolysis region,[29,34] and lightning and thunder shockwave production of HCN appears to be inadequate when observed production efficiencies in realistic laboratory discharge and shock-synthesis experiments are scaled to Jovian conditions.[10,23] We point out, however, that Bar-Nun and Podolak[24] and Podolak and Bar-Nun[25] continue to favor the thundershock hypothesis, and Fegley and Lodders[20] and Lodders and Fegley[21] still support a deep-atmospheric source.

The purpose of this paper is to use updated information on the kinetics of nitrogen species to reevaluate both the quenched chemistry (deep atmospheric source) and photochemistry (coupled $C_2H_2$-$NH_3$ photochemistry) hypotheses for the production of HCN on Jupiter. We use two different one-dimensional (1D) kinetic-transport numerical models for this investigation. Both are based on the Caltech/JPL KINETICS code,[43] which uses finite-difference techniques to solve the 1D continuity equations. The first model considers photochemical kinetics and molecular and eddy diffusion, and we apply that model to the stratosphere and upper troposphere of Jupiter. The second model considers thermochemical kinetics and eddy diffusion, and we apply that model to the deep Jovian troposphere. Previous investigators who have looked in detail at the possibility of HCN transport from the deep troposphere[44,20,21] have used time-constant arguments rather than full kinetic-transport models to predict the quenched HCN abundance in the upper troposphere. Recent suggested improvements to the time-constant arguments[45−48] have prompted us to reevaluate the thermochemical kinetics and quenching of the C-H-O system on Jupiter.[49] Our success at modeling the transition from thermochemical equilibrium to transport-induced quenching in that C-H-O system has led us to investigate nitrogen species thermochemistry and quenching for this study.

## 2. Nitrogen species kinetics

The full reaction list for our Jovian photochemical model includes 145 species and 1973 reactions. The H-C-O reactions are discussed elsewhere.[49,50] The kinetics of carbon-hydrogen-oxygen species has been well studied because of numerous combustion-chemistry applications[51−54] and terrestrial atmospheric-chemistry applications;[55−59] however, less information is available for reactions of nitrogen-bearing species, particularly in reducing environments. Some relevant experimental data are discussed in the above data compilations[51−59] and in numerous individual rate-coefficient studies. However, for our application, laboratory data must be supplemented by theory — both quantum chemistry and master-equation calculations — for many reactions of importance in our models. We therefore rely heavily on the theoretical calculations of Dean and Bozzelli[60] for the generation of our reaction mechanism. Several other investigations have also been useful for identifying important reactions and rate coefficients.[61−65] Our adopted reaction-rate coefficients are included in the full photochemical model output available in the Supplementary material.

We consider the kinetics of $H_2$, $H$, and 58 oxygen- and carbon-bearing species,[49,50] as well as $N$, $N_2$, $NH$, $NH_2$, $NH_3$, $NNH$, $N_2H_2$, $H_2NN$, $N_2H_3$, $N_2H_4$, $CN$, $HCN$, $H_2CN$, $CH_2NH$, $CH_3NH$, $CH_2NH_2$, $CH_3NH_2$ $CH_2CN$, $CH_3CN$, $C_3N$, $HC_3N$, $C_2H_2CN$, $C_2H_3CN$, $NO$, $NO_2$, $N_2O$, $HNO$, $HNO_2$, $NCO$, $HNCO$, $CH_3NO$, $PN$, and $NH_2PH_2$ in both our thermochemical and photochemical models. We also include several other phosphorus-bearing species, but a full discussion of the phosphorus kinetics and $NH_3$-$PH_3$ photochemical coupling is deferred to a later paper (see also[66,67]). We initially included $HNC$, $C_2N$, and $C_2N_2$ in the models, but these species were produced in trivial amounts in the photochemical models and had little effect on the kinetics of other constituents, so we dropped them from consideration. For our photochemical model, we also include $C_2H_4N$ (*i.e.*, $CH_2$=CH$\dot{N}$H, $CH_3\dot{C}$=NH, and/or $CH_3CH$=N· isomers), $C_2H_5N$ (*i.e.*, $CH_2$=CHNH$_2$ and $CH_3CH$=NH isomers), $C_2H_5\dot{N}$H, $C_2H_5NH_2$, $CH_3CH$=NNH$_2$, $CH_3CH$=NC$_2$H$_5$, and $CH_3CH$=NN=CHCH$_3$ based on laboratory photolysis investigations.[36−39] However, we omit these latter species from our deep-tropospheric thermochemical model due to a lack of information on their thermodynamic parameters — information that is needed to fully reverse the kinetic reactions through the principle of microscopic reversibility. Although these species are unlikely to be significant constituents in the deep troposphere of Jupiter, they are potentially important photochemical products of coupled $NH_3$-$C_2H_2$ photochemistry, as well as precursors to HCN formation in Jupiter's upper troposphere,[36−39] and must be included in the photochemical model.

Many of our reaction rate coefficients derive from Dean and Bozzelli,[60] who validate their proposed mechanism by comparing their model predictions with experimental results from several flame studies. The expressions provided for their individual reaction rate coefficients are valid for temperatures in the range of 600-2500 K. Those temperatures are appropriate for our deep-tropospheric thermochemical modeling but not for conditions in Jupiter's upper troposphere and stratosphere, where temperatures can drop to $\lesssim 110$ K at the tropopause.[68−71]  Therefore, although we generally adopt the Dean and Bozzelli expressions as given in their paper,[60] we check for pathological or inconsistent behavior at low temperatures and alter the expressions, as necessary. Some of the Dean and Bozzelli rate coefficients have also been replaced due to the availability of experimentally derived rate coefficients or due to inappropriate rate coefficients calculated for the reverse reaction (*i.e.*, those in excess of kinetic collision-rate considerations). Moreover, the Dean and Bozzelli[60] mechanism does not cover the full suite of nitriles, amines, hydrazones, and other complex nitrogen-bearing organics that are expected to form in coupled $NH_3$-$C_2H_2$ photochemistry.[36−39] Some important rate coefficients for the production and loss of these organo-nitrogen compounds could be found in the literature, but many could not. We therefore apply our reaction list to simulations of laboratory photolysis investigations to help constrain uncertain reaction rate coefficients and to test our overall mechanism.

The first such simulation we perform is for the photolysis of pure ammonia, as described in Groth *et al.*[72]  In this experiment, 37.5 torr of pure ammonia at room temperature is introduced to a cylindrical quartz cell 10-cm long and 5.5 cm in diameter and irradiated by 206.2-nm photons from an iodine lamp.[72]  The resulting photolysis product quantum yields are derived as a function of photons absorbed. To simulate this investigation, we use the KINETICS code[43] with our reaction list described above and apply it to a 1D "box" of the appropriate length (10 cm), with the appropriate 298 K, 37.5-torr $NH_3$ initial conditions. Since the $NH_3$ photolysis rate has a gradient within the cell under these conditions, we divide our box into a 21-segment grid. We assume a lamp flux at 206.2 nm at the front of our cell of $3 \times 10^{14}$ photons cm$^{-2}$ s$^{-1}$, although exact knowledge of the lamp flux is not necessary because the experimental results were reported in terms of quantum yields per quanta absorbed.[72]  After $1.03 \times 10^{19}$ photons absorbed, Groth *et al.*[72] find quantum yields for $N_2$, $H_2$, and $N_2H_4$ production of 0.163, 0.490, 0.0005, respectively. From our box-model simulations, we derive quantum yields of 0.162, 0.487, and 0.00115 for $N_2$, $H_2$, and $N_2H_4$, respectively, at a corresponding number of photons absorbed. As with the Groth *et al.* experiment[72], we find that the quantum yields of $H_2$ and $N_2$ level off after an initial rise, whereas the $N_2H_4$ quantum yield goes through an early maximum, followed by a drop off to low values. The results of our simulation are insensitive to reasonable assumptions about the diffusion coefficients within the cylinder or the lamp flux.

The photolysis of ammonia at 206.2 nm occurs exclusively through the $NH_3$ +

$h\nu \rightarrow NH_2(\tilde{X}\,^2B_1) + H$ pathway.[72] Hydrazine production and loss depends critically on these photolysis products, a fact that can explain the observed quantum-yield behavior of the $N_2H_4$. Hydrazine is produced in our model mainly from the termolecular reaction $2\,NH_2 + M \rightarrow N_2H_4 + M$, where M is any third body, and the dominant two loss processes (of roughly equal importance at later times) are $H + N_2H_4 \rightarrow N_2H_3 + H_2$ and $NH_2 + N_2H_4 \rightarrow N_2H_3 + NH_3$. When $NH_2$ is released from $NH_3$ photolysis, hydrazine is readily synthesized and continues to increase in concentration until enough $N_2H_4$, $NH_2$, and H is built up that loss processes can compete. At that point, the $N_2H_4$ reaches a constant concentration (*i.e.*, it is in steady state), while $N_2$ and $H_2$ production continues through the net equation $2\,NH_3 \rightarrow N_2 + 3H_2$, which goes through $N_2H_4$ and other $N_2H_x$ species as intermediates. For example, the dominant pathway for $N_2$ and $H_2$ production in our simulation is the following scheme:

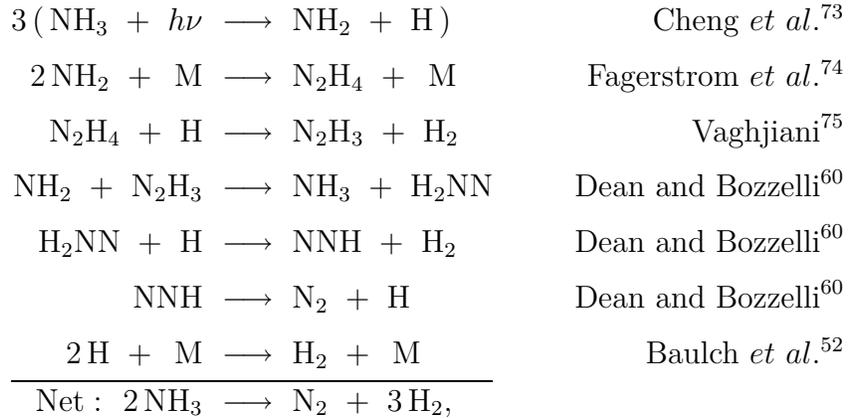

$$
\begin{array}{lr}
3\,(\,NH_3 \;+\; h\nu \;\longrightarrow\; NH_2 \;+\; H\,) & \text{Cheng } et\;al.^{73} \\
2\,NH_2 \;+\; M \;\longrightarrow\; N_2H_4 \;+\; M & \text{Fagerstrom } et\;al.^{74} \\
N_2H_4 \;+\; H \;\longrightarrow\; N_2H_3 \;+\; H_2 & \text{Vaghjiani}^{75} \\
NH_2 \;+\; N_2H_3 \;\longrightarrow\; NH_3 \;+\; H_2NN & \text{Dean and Bozzelli}^{60} \\
H_2NN \;+\; H \;\longrightarrow\; NNH \;+\; H_2 & \text{Dean and Bozzelli}^{60} \\
NNH \;\longrightarrow\; N_2 \;+\; H & \text{Dean and Bozzelli}^{60} \\
2\,H \;+\; M \;\longrightarrow\; H_2 \;+\; M & \text{Baulch } et\;al.^{52} \\
\hline
\text{Net}: \; 2\,NH_3 \;\longrightarrow\; N_2 \;+\; 3\,H_2, &
\end{array}
$$

where the reference at the end of each reaction represents the source of the rate coefficient or photolysis cross section. Note the importance of $N_2H_x$ intermediate species in this scheme (see also[76]); these species are also likely to be important in the kinetics of nitrogen species under combustion-chemistry conditions[60] and Jovian tropospheric conditions. Reactions involving $H_2NN$, a singlet biradical, are speculative at this point, but $H_2NN$ is expected to be a major product of the $NH_2 + NH_2$ reaction under combustion-chemistry conditions.[60] We strictly follow the Dean and Bozzelli theoretically derived rate coefficients[60] for the production and loss of this species and find that it can contribute to the conversion of $NH_3$ to $N_2$ under low-temperature $NH_3$ photolysis conditions.

A main secondary scheme in our simulation involves the $NH_2 + N_2H_4 \rightarrow NH_3$ + $N_2H_3$ abstraction reaction. Dean and Bozzelli[60] use a "Direct Hydrogen Transfer" (DHT) method to derive a rate coefficient of $6.1 \times 10^{-18}\,T^{1.94} \exp(-820/T)\,cm^3\ s^{-1}$ (for $T$ in K) for this reaction, which if extrapolated to 300 K would produce a value of $2.5 \times 10^{-14}$ cm$^3$ s$^{-1}$. In contrast, experimental data have been used to estimate a 300-K rate coefficient of $5 \times 10^{-13}$ cm$^3$ s$^{-1}$ for this reaction[77] — a value about 20 times larger than that derived theoretically.[60] We find that we get the best agreement with the Groth *et al.* experimentally derived $N_2$ and $H_2$ quantum yields[72] if we adopt a value that lies in between the experimental[77] and theoretical[60] values, *i.e.*, if we adopt a rate coefficient of $4.3 \times 10^{-17} T^{1.94} \exp(-820/T)$ cm$^3$ s$^{-1}$ for this reaction (seven times the Dean and Bozzelli[60] expression). We use this expression throughout our subsequent modeling.

Although our modeled quantum yields (and their time variation) for $N_2$ and $H_2$ are in excellent agreement with experimental results,[72] our quantum yields for $N_2H_4$ do not exactly match the experiments. Our $N_2H_4$ quantum yield goes through a maximum at slightly earlier times (albeit at a similar peak magnitude of $\sim$0.03-0.04) and falls off more quickly initially than was observed,[72] then more slowly at later times. Moreover, our $N_2H_4$ concentration reaches a steady state (with a quantum yield that therefore linearly decreases with photons absorbed), whereas the $N_2H_4$ concentration in the experiments apparently decreases after an early maximum before possibly reaching a low constant value (see Fig. 1 in Groth *et al.*[72]). Photolysis of $N_2H_4$ is included in our model, and although occurring, the $N_2H_4$ loss due to photolysis cannot compete with abstraction by hydrogen atoms and $NH_2$. It is unclear what the additional loss process might be. Despite this slight quantitative inconsistency with the $N_2H_4$ behavior, we have chosen not to tweak the reaction rate coefficients any further, as the dominant reactions (except for the one for $NH_2 + N_2H_4$, which we modified) all have literature-derived values. Keep in mind, however, that our mechanism may slightly overpredict the net production of $N_2H_4$ under these conditions.

The second simulation we perform is the investigation of the photochemical coupling of ammonia ($NH_3$) and acetylene ($C_2H_2$) in the presence of $H_2$.[35−39] As has been discussed in detail in these investigations, several complex organo-nitrogen compounds are produced when $H_2$-$NH_3$-$C_2H_2$ mixtures are irradiated by 184.9-nm and 206.2-nm photons. The rate coefficients for the production and loss of these compounds are generally not available in the literature. We therefore simulate the conditions in the Keane *et al.* experiments[39] and compare models to experimental results in order to help constrain the relevant kinetics.

The specific experiment we simulate is the irradiation of a mixture of 600 torr $H_2$, 40 torr $NH_3$, and 5 torr $C_2H_2$ at 296 K by 206.2-nm ultraviolet photons. Keane *et* *al.*[39] present quantum yields resulting from that investigation; however, we utilize more detailed information on this experiment, including the time variation of the photoproducts, from the thesis and laboratory notebooks of T. C. Keane.[38] During the experiment, quartz cells of 2.5-cm diameter and 10-cm length were filled with the above gas mixture, and an iodine lamp was used to irradiate the cell with 206.2 nm photons for various lengths of time. Ammonia actinometry[76] was used to determine that $4.438 \times 10^{15}$ photons per second were entering the cell, for a corresponding 206.2-nm flux of $9.04 \times 10^{14}$ photons cm$^{-2}$ s$^{-1}$. The composition and abundance of the photoproducts were measured by 500 MHz NMR spectroscopy, and full details of the experimental and analysis procedure can be found in the original reports.[38,39]

We use KINETICS[43] to simulate this experiment in a similar manner as for the pure ammonia photolysis experiment.[72] We start with a one-dimensional 10-cm box (subdivided into a six-segment grid) filled with the appropriate 600/40/5 torr mix of $H_2$/$NH_3$/$C_2H_2$ irradiated by a 206.2-nm flux of $9.04 \times 10^{14}$ photons cm$^{-2}$ s$^{-1}$, and use the KINETICS model with our full 1973-reaction list to solve for the time variation in the abundances of carbon-, nitrogen-, and hydrogen-bearing species. Our model results are compared with the experimental results[38] in Fig. 1.

The dominant nitrogen-bearing products in our model are $N_2$ (not investigated by Keane 1995), acetaldazine ($CH_3CH=N-N=CHCH_3$), ethylamine ($C_2H_5NH_2$), and *N*-ethylethylideneimine ($CH_3CH=NC_2H_5$), with lesser amounts of the other species shown in Fig. 1. Almost no quantitative kinetic information exists for these species. Kaye and Strobel[35], Ferris and Ishikawa[37], and Keane[38] all propose that the critical first step in the coupled photochemistry of $NH_3$-$C_2H_2$ is the sequence $NH_3 + h\nu \rightarrow NH_2 + H$ and $C_2H_2 + H + M \rightarrow C_2H_3 + M$, followed by $NH_2 + C_2H_3 + M \rightarrow C_2H_5N + M$. However, the identity and fate of the $C_2H_5N$ isomer, and the resulting production sequences for the different complex organo-nitrogen species differ between these investigations. Kaye and Strobel[35] do not attempt to distinguish the main $C_2H_5N$ isomer, although they do note that aziridine, a cyclic $C_2H_5N$ isomer, is known to yield HCN upon photolysis,[78] and they suggest that the other isomers would, as well. Kaye and Strobel[35] stop their mechanism at the $C_2H_5N$ photolysis stage and do not further investigate the production of complex nitrogen-bearing organics. Ferris and Ishikawa[37] and Keane[38] suggest that vinylamine ($C_2H_3NH_2$) is formed initially, followed by isomerization to ethylideneimine ($CH_3CH=NH$), a process that is common with enamines.[79−81] Keane[38] goes on to propose that the key to the formation of HCN, as well as some of the complex species is the ethylideneiminyl radical ($CH_3CH=N\cdot$), which can react with itself to form acetaldazine, or thermally decompose (or photolyze) to form HCN or $CH_3CN$. Although we do not explicitly distinguish between $C_2H_5N$ and $C_2H_4N$ isomers in our model, our reaction list implicitly follows the main $CH_3CH=NH$ and $CH_3CH=N\cdot$ pathways suggested by Keane.[38] However, some of our production mechanisms for the complex species diverge from those suggested by the earlier investigations.[37,38]

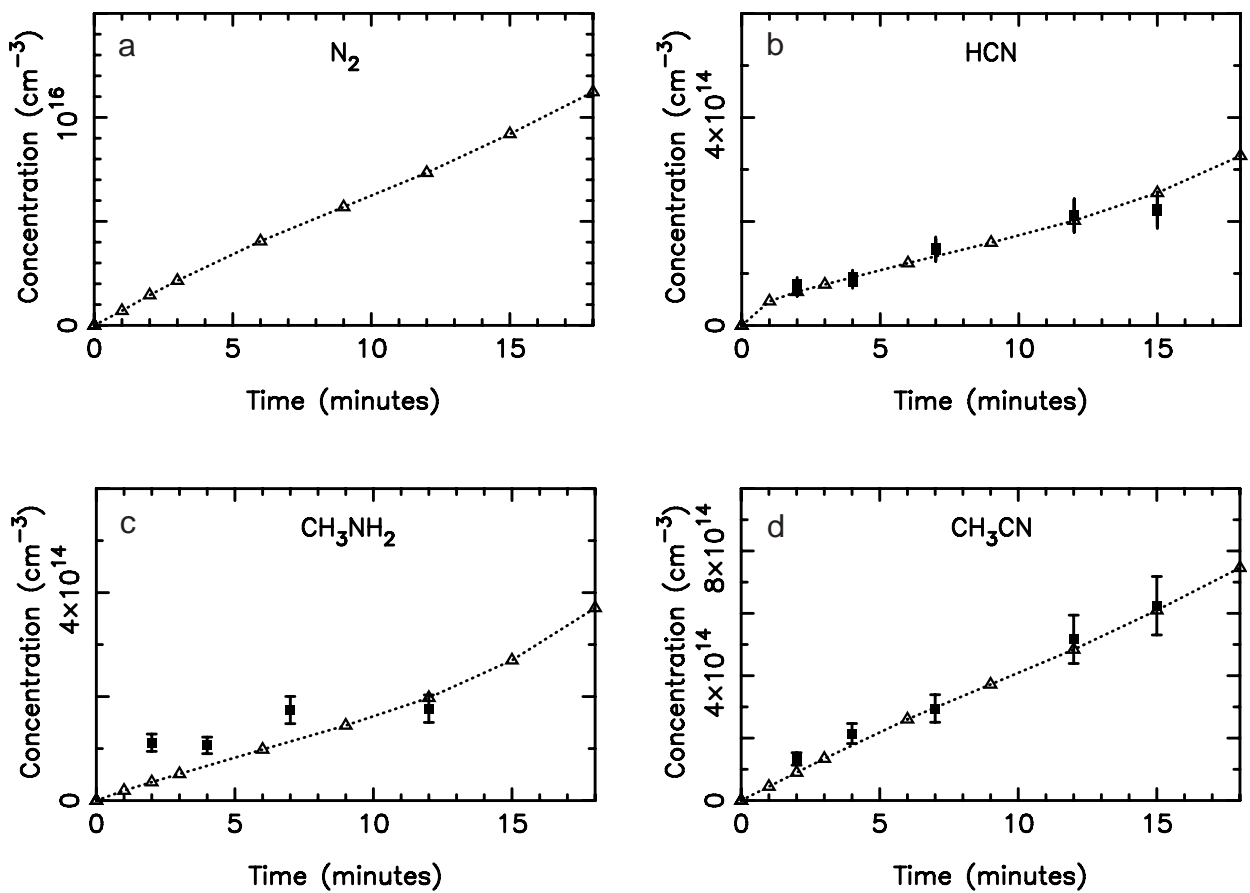

FIGURE 1. Concentrations of (a) molecular nitrogen, (b) hydrogen cyanide, (c) methy-lamine, (d) acetonitrile, (e) ethylamine, (f) acetaldehyde hydrazone, (g) *N*-ethylethylideneimine, and (h) acetaldazine as a function of time in the photolysis cell. The dotted lines and open triangles correspond to the model results, whereas the solid squares correspond to the experimental results.[38] Note the change in the ordinate range for the different species. Reported measurement errors are 10-15% on species abundances, mostly due to the NMR technique.[38]

As an example, the dominant acetaldazine (see Fig. 1h) formation mechanism in our model is $2\,C_2H_4N + M \rightarrow CH_3CH=NN=CHCH_3$, as suggested by Keane[38] and

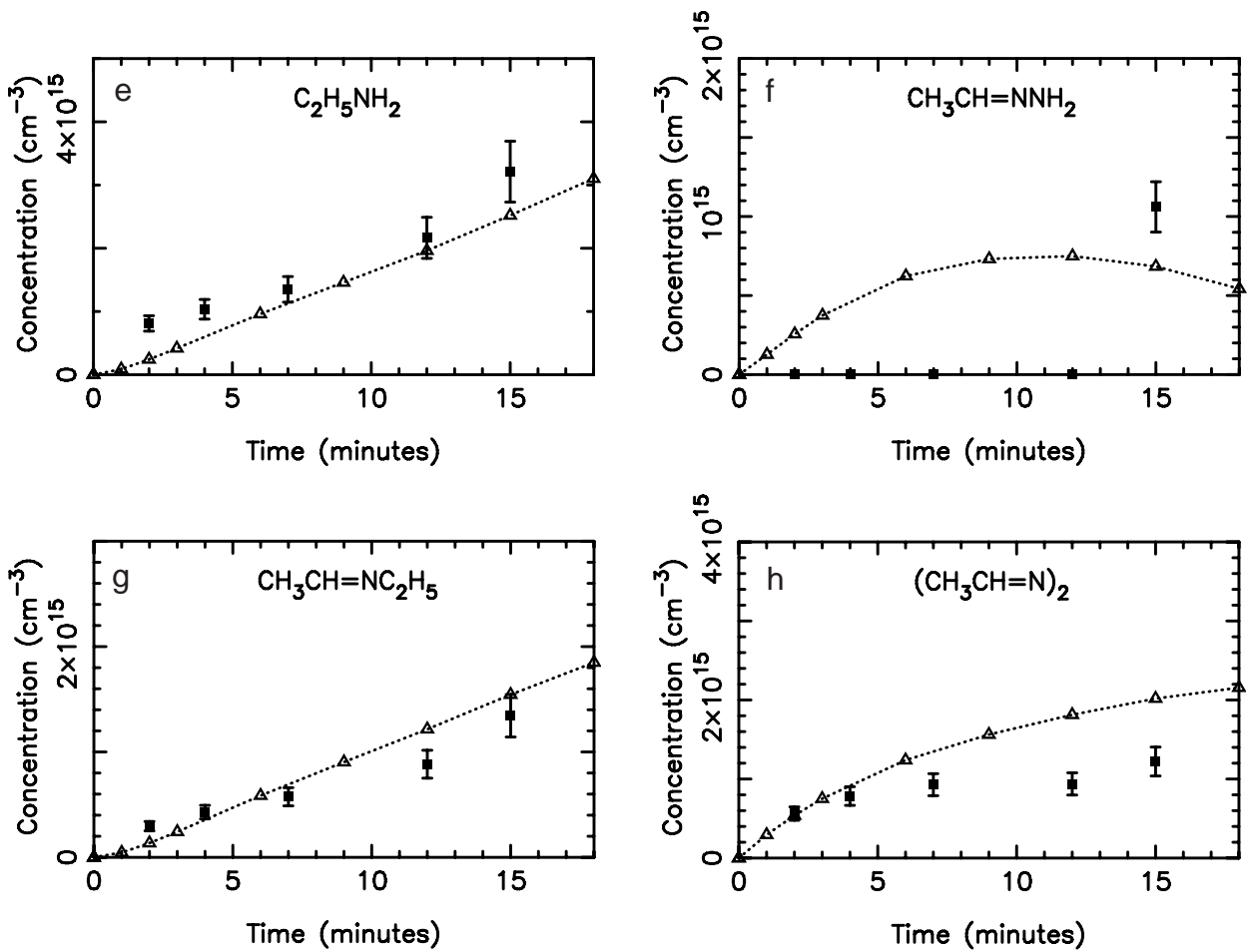

**FIGURE 1.** *(cont.)*

Ferris *et al.*,[37] where their expected $C_2H_4$ isomer is $CH_3CH=N\cdot$. However, the main non-recycling production pathway for $C_2H_4N$ in our model is the reaction $C_2H_2 + NH_2 + M \rightarrow C_2H_4N + M$, followed very closely in terms of importance by hydrogen abstraction from $C_2H_5N$ by both H and $NH_2$, whereas only the H-atom abstraction pathway $C_2H_5N$ (*i.e.*, as $CH_3CH=NH$) + H $\rightarrow C_2H_4N + H_2$ is considered in the earlier investigations.[37,38] Note that the $C_2H_2 + NH_2 + M$ reaction has been studied experimentally[82–85] and theoretically,[86] and we adopt a rate coefficient for this reaction of $k_0 \approx 1 \times 10^{-26}$ cm$^6$ s$^{-1}$ and $k_\infty = 1.3 \times 10^{-19} T^{2.03} \exp(-1300/T)$ cm$^3$ s$^{-1}$, where $T$ is the temperature in kelvins. Photolysis of acetaldazine helps recycle $C_2H_4N$ in our model and is the main effective loss process of acetaldazine.

Ethylamine (Fig. 1e) is produced in our model through the reaction $NH_2 + C_2H_5 + M \rightarrow C_2H_5NH_2 + M$, with loss due to photolysis and abstraction by H atoms to form $C_2H_5\dot{N}H$. These pathways are consistent with previous suggestions.[38] Although the rate coefficient for the reaction of $C_2H_5$ with $NH_2$ has been measured,[87] the relative roles of addition and disproportionation are not clear, and the assumed relative rates of the different product pathways for this reaction can affect our results. To prevent large quantities of ethylamine from building up in our cell, we have assumed that $NH_2$ + $C_2H_5 \rightarrow NH_3 + C_2H_4$ is about three times faster than $NH_2 + C_2H_5 \rightarrow C_2H_5NH_2$. This factor-of-three value may be a overestimate, as can be seen from the fact that we slightly underestimate the ethylamine abundance shown in Fig. 1e. A thorough discussion of the photolysis of ethylamine can be found elsewhere.[38]

Keane[38] suggests that $N$-ethylethylideneimine (Fig. 1g) is produced through the nucleophilic addition/elimination reaction $CH_3CH{=}NH + C_2H_5NH_2 \rightarrow CH_3CH{=}N$-$C_2H_5 + NH_3$. However, this reaction may have a significant activation energy barrier in the gas phase, given that the reactants are molecules and not radicals. We instead suggest that $CH_3CH{=}N$-$C_2H_5$ formation might occur through reactions such as (1) $CH_3CH{=}N{\cdot} + C_2H_5 \rightarrow CH_3CH{=}N$-$C_2H_5$, (2) $C_2H_5N + C_2H_5\dot{N}H \rightarrow CH_3CH{=}N$-$C_2H_5 + NH_2$, (3) $CH_3CH{=}N{\cdot} + C_2H_6 \rightarrow CH_3CH{=}N$-$C_2H_5 + H$, (4) $CH_3CH{=}N{\cdot} + C_2H_5NH_2 \rightarrow CH_3CH{=}N$-$C_2H_5 + NH_2$, (5) $C_2H_5\dot{N}H + C_2H_5 \rightarrow CH_3CH{=}N$-$C_2H_5 + H_2$, (6) $C_2H_5N + C_2H_5 \rightarrow CH_3CH{=}N$-$C_2H_5 + H$, or (7) $C_2H_5NH_2 + C_2H_3 \rightarrow CH_3CH{=}N$-$C_2H_5 + H$. Without thermodynamic parameters for $N$-ethylethylideneimine and some of these other species, the exothermicity and likely activation energies of these reactions cannot be determined. In our model, the first of our suggested reactions dominates $N$-ethylethylideneimine production, but other potential pathways are also included in the model; see the Supplementary material for our estimates of these reaction rate coefficients. $N$-ethylethylideneimine is destroyed in our model predominantly through reactions with H (to form either $CH_3CH{=}N{\cdot} + C_2H_6$ or $C_2H_5N + C_2H_5$) and with $CH_3CH{=}N{\cdot}$ (to form acetaldazine + $C_2H_5$). The production and loss mechanisms for $N$-ethylethylideneimine remain considerably uncertain. In particular, given the rapid observed rate of the gas-phase reaction of $CH_3CHO$ and $N_2H_4$ to form $CH_3CH{=}N$-$NH_2$ and other products,[38] it is possible that $CH_3CH{=}NH$ participates in nucleophilic addition/elimination reactions without much of an energy barrier in the gas phase, as originally proposed by Keane.[38]

Acetaldehyde hydrazone exhibits interesting behavior in Keane's experimental data (see Fig. 1f). At room temperature, the $CH_3CH{=}NNH_2$ abundance is negligible until irradiation times of 15 minutes or longer, at which point it jumps up to become a major product of coupled $C_2H_2$-$NH_3$ photochemistry. The reason for this late production is unclear, and the model does not reproduce this behavior. One possibility is that slow photolysis of one of the more abundant photoproducts such as $C_2H_5NH_2$, $CH_3CH=NC_2H_5$, or $CH_3CH=NN=CHCH_3$ contributes, but the profiles of these species show no evidence for a sudden, significant loss at late times. Atomic hydrogen begins to build up in our model at later times as more and more of the $C_2H_2$ is destroyed, and a second possibility is that hydrogen abstraction commences as a significant loss process for some of our species as H atoms build up in the cell, with a corresponding significant increase in the production rate for $CH_3CH=NNH_2$. However, again, there are no signs of the corresponding reduction of any of the other observed species, and we are unable to find a combination of reactions that would reproduce this behavior. The late production of acetaldehyde hydrazone remains a mystery. In our model, $CH_3CH=NNH_2$ is produced predominantly through the addition reactions $NH_2 + C_2H_4N \rightarrow CH_3CH=NNH_2$ and $H_2NN + C_2H_4 \rightarrow CH_3CH=NNH_2$ and is lost mainly through $H + CH_3CH=NNH_2 \rightarrow NH_2 + C_2H_5N$, but given our poor comparisons with the experimental data,[38] we have no confidence in our adopted kinetics for $CH_3CH=NNH_2$.

Kinetic information for some of the simpler species like HCN, $CH_3CN$, and $CH_3NH_2$ exists in the literature, but that does not mean the production and loss of these species under the Keane[38] experimental conditions are well understood. Methylamine (Fig. 1c) is produced in our model largely from $NH_2 + CH_3 + M \rightarrow CH_3NH_2 + M$, for which rate-coefficient information is available.[88] Photolysis dominates the loss of $CH_3NH_2$, a process that has been well studied.[89−93] Acetonitrile (Fig. 1d) has four main production mechanisms in our model, all involving Keane's key intermediate, the $CH_3CH=N\cdot$ radical.[38] In order of decreasing importance, these are (1) $2CH_3CH=N\cdot \rightarrow CH_3CN + C_2H_5N$, (2) $CH_3CH=N\cdot + C_2H_5 \rightarrow CH_3CN + C_2H_6$, (3) $CH_3CH=N\cdot + H \rightarrow CH_3CN + H_2$, and (4) $CH_3CH=N\cdot + NH_2 \rightarrow CH_3CN + NH_3$, for which there are no literature values for rate coefficients. In our model, $CH_3CN$ is destroyed largely through the abstraction reactions $CH_3CN + CH_3 \rightarrow CH_2CN + CH_4$ and $CH_3CN + H \rightarrow CH_2CN + H_2$, as well as through $CH_3CN + H \rightarrow HCN + CH_3$. Rate coefficients for the two reactions of H with $CH_3CN$ have been reported.[63] The reaction $CH_3CN + CH_3 \rightarrow CH_2CN + CH_4$ is exothermic under our conditions but likely possesses a significant activation barrier. We estimate a rate coefficient of $1.0 \times 10^{-12} \exp(-3000/T)$ (for $T$ in K) for this reaction, based in part on analogy with $H + CH_3CN$.

Hydrogen cyanide (Fig. 1b) is produced in multiple ways under the conditions of this experiment, including from photolysis of many of the above species.[38] Due to the rapid synthesis of $C_2H_4N$ and H radicals with our proposed mechanism, the dominant production pathway in our model is the speculative reaction $C_2H_4N + H \rightarrow HCN + CH_4$. This reaction is likely to be highly exothermic; however, other product pathways may compete, including formation of the $C_2H_5N$ adduct, formation of $CH_3CN + H_2$, and, less likely because of the amount of rearrangement involved and the necessity of breaking the strong C=N bond, the formation of $NH_3 + C_2H_2$. We have adopted a rate coefficient of $4 \times 10^{-11}$ $cm^3$ $s^{-1}$ for the $C_2H_4N + H \rightarrow HCN + CH_4$ reaction. Other significant production mechanisms in our model involve $H_2CN$, either through self reaction,[94] reaction with H,[95] or reaction with $NH_2$.[60] Hydrogen cyanide is lost mainly through the reaction $HCN + C_2H_3 \rightarrow C_2H_3CN + H$.[96]

The acrylonitrile ($C_2H_3CN$, not shown in Fig. 1) that forms in the latter reaction also builds up in the cell in our simulation, for an overall abundance that lies in between that of acetonitrile and ethylamine. Keane[38] finds no evidence for acrylonitrile in the NMR spectra, although he does note that a liquid polymer forms on the window of the cell during the photolysis experiment. It is possible that the $C_2H_3CN$ polymerizes or that we have neglected some other significant loss process for this molecule. The dominant loss mechanism currently in our model is the reverse of the production reaction, *i.e.*, $C_2H_3CN + H \rightarrow HCN + C_2H_3$.

Two other species that form in noticeable quantities in our model but are not detected by Keane[38] are methanimine ($CH_2NH$), whose concentration reaches a peak value of $5.8 \times 10^{14}$ $cm^{-3}$ after 21 minutes before slowly dropping off with time, and $C_2H_5N$, whose concentration peaks at $4.6 \times 10^{14}$ $cm^{-3}$ after 15 minutes before slowly dropping off with time. Methanimine is produced in our model mainly from reaction of $H_2CN$ with $C_2H_3$ and $C_2H_5$ radicals and is lost from abstraction by $C_2H_3$ to form $H_2CN + C_2H_4$. Because of its suspected importance in their overall reaction mechanism, Keane[38] searched specifically for a signature that could be caused by the $C_2H_5N$ isomer ethanimine ($CH_3CH=NH$), as well as attempted to synthesize and isolate this imine, but both attempts were unsuccessful. The primary production mechanism for forming $C_2H_5N$ in our model, and indeed one of the top three mechanisms for forming the C-N bond in our simulation, is the reaction originally proposed by Kaye and Strobel:[35] $NH_2 + C_2H_3 + M \rightarrow C_2H_5N + M$. Reaction of $CH_3CH=NNH_2$ with H to form $C_2H_5N + NH_2$ contributes at later times. The primary loss processes for $C_2H_5N$ in our model include abstraction by H or $NH_2$ to form $C_2H_4N$ and $H_2$ or $NH_3$, and reaction with $C_2H_5\dot{N}H$ to form $CH_3CH=N-C_2H_5 + NH_2$. Aside from the $C_2H_3 + NH_2 \xrightarrow{M} C_2H_5N$ reaction proposed earlier,[35] the two other key reactions leading to the formation of the C-N bond in our model are $NH_2 + C_2H_2 \xrightarrow{M} C_2H_4N$ and $NH_2 + C_2H_5 \xrightarrow{M} C_2H_5NH_2$ (see above).

Keane[38] does not track the time dependence of the $N_2H_4$ abundance in their experiment. However, a quantum yield is reported[39] for $N_2H_4$ formation of 0.007 from the first few minutes of the irradiation (further details are not specified). We obtain that $N_2H_4$ quantum yield after 7 minutes in our model.

Although our model results compare well with the time variation observed for many of the species in the Keane[38] experiment (see Fig. 1), problems do exist, especially for $CH_3CH=NNH_2$ (Fig. 1f), and to a lesser extent for $CH_3CH=NN=CHCH_3$ (Fig. 1h). The actual mechanism involved is likely far more complex than our limited reaction list can attempt to reproduce, but without further information on the thermodynamic properties of the key species and on the rate coefficients for individual reactions, we are unlikely to implement meaningful improvements to the proposed mechanism. Despite the incomplete and cursory nature of the proposed mechanism, we at least now have in place a reaction list that includes estimates for the production and loss of the major species involved in the coupled chemistry of $C_2H_2$-$NH_2$-$H_2$ mixtures, and we can test the effectiveness of coupled $NH_3$-$C_2H_2$ photochemistry under Jovian conditions. We can test, in particular, whether the addition of these various complex nitrogen-bearing species contributes to the formation of HCN on Jupiter.

## 3. Photochemical model

The photochemistry of ammonia on Jupiter was first investigated qualitatively by Wildt[97] and quantitatively by Cadle[98] and McNesby;[99] the latter two authors both discussed the likelihood of carbon-nitrogen coupling. Strobel[100] was the first to develop a realistic model that explained the continuing presence of $NH_3$ on Jupiter, through a nitrogen cycle in which convection allows photochemical products like $N_2H_4$ to be transported to deeper, hotter levels of the troposphere, where they can be converted back into $NH_3$. Strobel[100] suggested that slow vertical mixing above the ammonia clouds, combined with efficient $NH_3$ photolysis and $N_2H_4$ production, would limit the abundance of $NH_3$ in Jupiter's stratosphere and thus inhibit carbon-nitrogen coupling; he was also the first to suggest that condensed $N_2H_4$ could be a major component of the Jovian upper tropospheric and lower stratospheric haze. Atreya *et al.*[101] further refined these earlier models and considered the possible effects of hydrazine supersaturation on the distribution of nitrogen-bearing species, Kuhn *et al.*[29] examined $CH_4$-$NH_3$ coupled photochemistry through the production of methylamine, and Strobel[102] examined $NH_3$-$PH_3$ photochemical coupling. Kaye and Strobel[34] refuted the suggestion[29] that methylamine production would be significant on Jupiter and demonstrated that $CH_3NH_2$ and HCN production would be greatly inhibited even if the catalytic destruction of $CH_4$ through $C_2H_2$ photolysis products were included in the model. Kaye and Strobel[35] suggested instead that coupled $NH_3$-$C_2H_2$ photochemistry could lead to the production of HCN and other carbon-nitrogen species on Jupiter.

After the seminal works of Kaye and Strobel,[34,35,66,67] there was a hiatus in Jovian tropospheric photochemical modeling until observations became advanced enough to provide constraints on the vertical and horizontal profiles of ammonia and other tropospheric species. The first to exploit these advances were Edgington *et al.*,[103,104] who created a photochemical model to investigate the latitude and altitude variation of ammonia and phosphine from ultraviolet Hubble Space Telescope observations. However, the Edgington *et al.* models were based on the earlier models[101,105] that did not include the coupled $C_2H_2$-$NH_3$ photochemistry suggested by theoretical[35] and experimental[36−39] studies. The models we develop for this paper represent the first attempt to include the photochemical production of the complex organo-nitrogen species observed in these $NH_3$-$C_2H_2$ photolysis experiments. Advances in our knowledge of the vertical distribution of temperatures, stratospheric hydrocarbons, and tropospheric constituents on Jupiter[106,107] greatly aid our current investigation. Details concerning the photochemical model and our assumptions and inputs to that model are described below.

## 3.1 Photochemical model inputs

Our photochemical model is designed to represent global-average conditions on Jupiter. We adopt a temperature profile for the upper troposphere and middle atmosphere as described by Moses *et al.*[50] The temperature profile at pressures less than 0.001 mbar derives from the Galileo probe ASI data,[69] whereas the profile at pressures greater than 1 mbar derives largely from global-average Infrared Space Observatory observations;[70] in between these pressure regions, a roughly isothermal atmosphere is assumed. The model extends from 6.7 bar to $2.3 \times 10^{-8}$ mbar, in a grid of 111 pressure levels, with a vertical resolution of at least three levels per atmospheric scale height. This extensive vertical range allows us to encompass not only the $NH_3$ photolysis region in the upper troposphere, but the methane photolysis region in the upper stratosphere. Low-to-average solar ultraviolet flux values are adopted,[50] and Jupiter's orbital distance is fixed at 5.2 AU from the Sun. As with the model described in Section 2, our reaction list contains 145 C-H-O-N-P species and 1973 reactions. Diurnally averaged quantities are considered, and we run the model until steady-state conditions are achieved. Condensation of $NH_3$, $N_2H_4$, $P_2H_4$, HCN, $CH_3CN$, and $H_2O$ (the latter from external sources) are included in a manner described elsewhere.[108] Hydrogen sulfide ($H_2S$) and the tropospheric water and $NH_4SH$ clouds, with their potential effects on ammonia and other constituents,[107,109] are not considered.

At the lower boundary of the model, we assume a He mole fraction of 0.136,[110,111] a $CH_4$ mole fraction of $2.04 \times 10^{-3}$,[112] a CO mole fraction of $1.0 \times 10^{-9}$,[48] a $PH_3$ mole fraction of $7 \times 10^{-7}$,[113] and an $NH_3$ mole fraction of either $1 \times 10^{-4}$,[114] $3 \times 10^{-4}$,[115,116] or $5.72 \times 10^{-4}$.[112] The three values for the bottom $NH_3$ boundary condition were chosen to reflect the differences between the Jovian belt regions, for which $NH_3$ is apparently dynamically depleted at the few-bar level, and zone regions, for which $NH_3$ might achieve its deep, well-mixed value at the few-bar level.[107,109,112,115,117−118] All other species are assumed to have a zero concentration gradient at the lower boundary such that the species are transported through the lower boundary at a maximum possible rate. Zero flux boundary conditions are adopted at the upper boundary for all species except H, $H_2O$, CO, and $CO_2$. Atomic hydrogen, which is produced in the thermosphere, is assumed to have a downward flux of $8.0 \times 10^8$ atoms $cm^{-2}$ $s^{-1}$ at the top boundary.[50] We also assume that $H_2O$, CO, and $CO_2$ from external sources are entering at the top of the atmosphere with fluxes of, respectively, $4 \times 10^4$, $1 \times 10^6$, and $1 \times 10^4$ $cm^{-2}$ $s^{-1}$.[8,48,50,119]

Both eddy and molecular diffusion are considered in the transport terms of the continuity equations. Our adopted molecular diffusion coefficients are described elsewhere.[120] In 1D models, the "eddy" diffusion coefficient $K_{zz}$ provides a convenient means by which to parameterize vertical motions of all scales in the atmosphere. The stratospheric values for $K_{zz}$ are taken from Moses *et al.*,[50] who provide a full discussion of the various observations that were used to help constrain the eddy-diffusion-coefficient profile. All these observational constraints have model dependencies that are difficult to characterize quantitatively. Near the methane homopause region in the upper stratosphere, $K_{zz}$ is constrained from observations of the methane density profile from spacecraft ultraviolet occultation observations,[121,122] from ground-based near-infrared stellar occultation observations,[123] and from ultraviolet airglow observations of the 121.6-nm

H Lyman $\alpha$ line emission and the 58.4-nm He line emission.[124,125] In the middle strato-sphere, near the few tenths of a millibar level, $K_{zz}$ is constrained from observations of the evolution of the Shoemaker-Levy 9 vapor deposited during the plume splashback phase of the impacts.[5,8,119] In the lower stratosphere ($\sim$1-100 mbar), $K_{zz}$ is constrained from mid-infrared ethane observations.[50] The minimum value of $K_{zz} \approx$ 300-1500 cm$^2$ s$^{-1}$ in the upper troposphere derives from CO observations[48] (see also the inferences from infrared and ultraviolet observations of PH$_3$ and NH$_3$ [103,104,126−128]). This minimum value affects the profiles of the species that have a production source at high altitudes; a low value (*i.e.*, a stagnant lower stratosphere) allows a greater build up of the column abundance of photochemically stable species like CO and C$_2$H$_6$ in the stratosphere.[129] Due to the numerous observational constraints on stratospheric $K_{zz}$ values, we keep our stratospheric $K_{zz}$ profile fixed but allow the tropospheric value to vary as a free parameter.

Tropospheric $K_{zz}$ values are difficult to constrain by remote-sensing observations. Within and below the water clouds, convective motions are expected to result in high effective eddy diffusion coefficients of order $K_{zz} \approx 10^8$-$10^9$ cm$^2$ s$^{-1}$.[130] However, when ra-diative processes become important and the temperature profile ceases to become purely convective (*i.e.*, in the upper troposphere above the clouds), vertical mixing is expected to be reduced. The PH$_3$ and NH$_3$ vertical profiles derived from infrared and ultraviolet observations have been used in this region, usually in combination with photochemistry or diffusion models, to constrain $K_{zz}$ in the upper troposphere, although the results are very model dependent.[103,104,126,128] Moreover, these observations demonstrate that the PH$_3$ and NH$_3$ profiles, and the resulting $K_{zz}$ inferences, vary with latitude, whereas we are attempting to construct a global-average model. As we will show, our results have some sensitivity to the adopted tropospheric $K_{zz}$ values: high values allow for rapid removal of photochemically generated species into the deep troposphere. Based on Edg-ington *et al.*,[104] we adopt a $K_{zz}$ profile that varies slowly and linearly with pressure in the troposphere with a value of $\sim$1-2 $\times$ 10$^4$ cm$^2$ s$^{-1}$ between $\sim$180 mbar and 1 bar, and we test the sensitivity of the results to the tropospheric eddy diffusion coefficient by simply multiplying this linear $K_{zz}$ profile by a constant value.

Rayleigh scattering of H$_2$, He, and CH$_4$ has been included in the model. Aerosol opacity in the 150-230 nm wavelength region can also influence our model results, through shielding of PH$_3$, NH$_3$, and other molecules from photolysis. West *et al.*[131] review our current state of knowledge of the Jovian cloud and haze properties. Un-fortunately, there appears to be little consensus regarding the details of the optical and physical properties and structure of the upper tropospheric and lower stratospheric clouds and hazes on Jupiter, as different groups, using different data sets and analysis procedures, derive different results. Uniquely deriving these properties is difficult due to the large number of unknown parameters that can strongly affect the results, and the problem is exacerbated by horizontal variations in these properties across Jupiter. Because of a lack of reliable, detailed information on the aerosol properties in this wavelength region,[103,104,132,133] we simplify the problem by including aerosol absorption only and neglect aerosol scattering. For our nominal model, we assume an optically thick haze in the 300-700 mbar region with vertical optical depth 2.7 at 150-230 nm, presumed to be the $NH_3$ cloud itself, and an optically thin haze in the 10-150 mbar region of vertical optical depth 0.14 (see West *et al.*,[131] Sromovsky and Fry,[134] and references therein for further details and comparisons with other models). We test the sensitivity of these results to the assumed optical thickness of the 300-700 mbar haze. Fortunately, condensation itself (and the saturation vapor pressure variation with temperature) has the largest effect on the vertical profile of $NH_3$ in the upper troposphere, and our results regarding the coupled $NH_3$-$C_2H_2$ photochemistry are relatively insensitive to our aerosol-opacity assumptions.

## 3.2 Photochemical model results

Our model results in terms of the mole-fraction profiles for several important nitrogen-bearing species in our nominal model are shown in Fig. 2. Also shown in the figure is the $C_2H_2$ profile derived from the model. Note that although $C_2H_2$ is abundant in the stratosphere, its mole fraction is expected to fall off significantly with decreasing altitude because of reaction with atomic H, photolysis, and other photochemical loss processes. Our model atmosphere contains only 0.11 ppb of $C_2H_2$ at the tropopause (140 mbar). Similarly, the $NH_3$ mole fraction in the tropopause region is also significantly reduced by condensation, photolysis, and other loss processes. Photochemical coupling of $C_2H_2$-$NH_3$ is therefore greatly inhibited compared to static photolysis experiments[38] described above, in which the initial mole fractions of $NH_3$ and $C_2H_2$ were 6.2% and 0.775%. Some $NH_3$-$C_2H_2$ coupling does occur, however, in our Jovian photochemical model, resulting in the production of small amounts of HCN, $CH_3CN$, $C_2H_5NH_2$, and the other complex organo-nitrogen species seen in the photolysis experiments.[36−39] Hydrogen cyanide is the dominant end product of this coupled chemistry, but its predicted abundance is well below the observational upper limits.[13−15] Our full photochemical model output, including species abundances, reaction rate coefficients, photolysis rates, production and loss rates, and chemical loss time scales can be found in the Supplementary material.

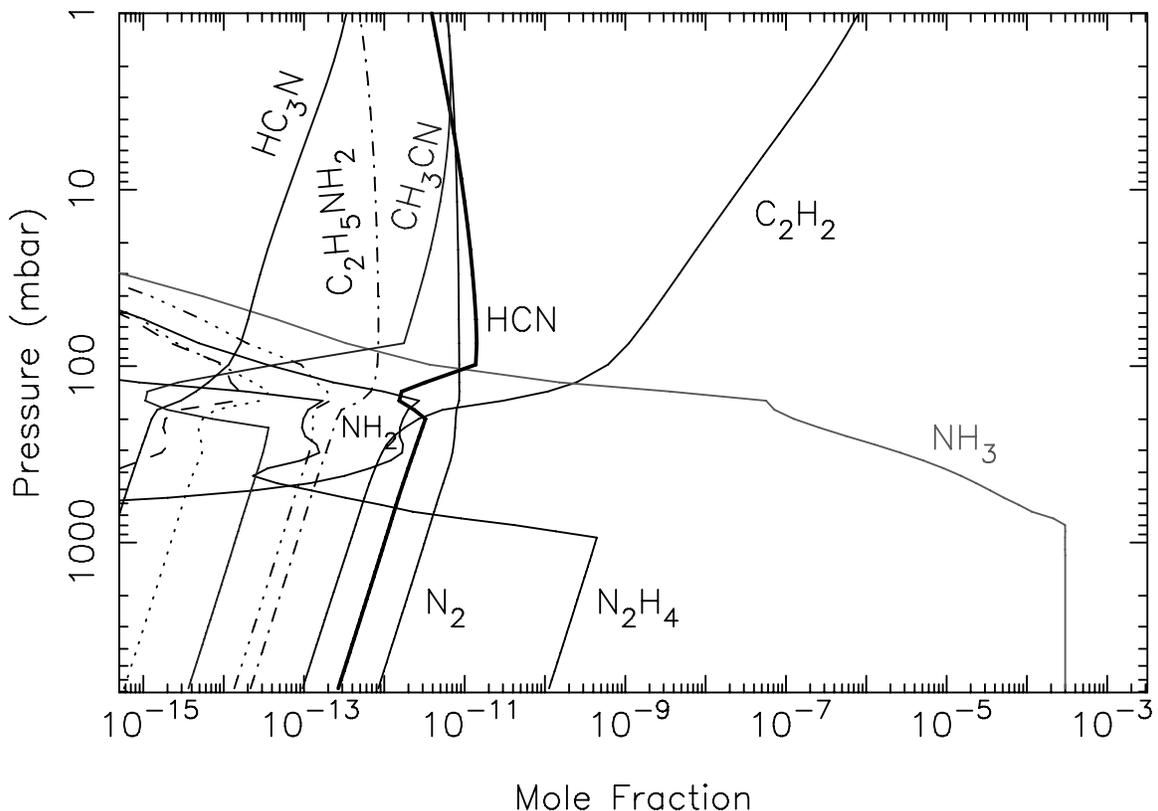

FIGURE 2. Mole-fraction profiles for several nitrogen-bearing species in our nominal photochemical model (as labeled). The triple-dot-dashed profile represents $CH_2NH$, the dotted profile represents $CH_3CH{=}NH$, and the dashed profile represents $CH_3NH_2$. The acetylene profile is included for comparison; $C_2H_2$ is relatively abundant in the stratosphere but its mole fraction falls off rapidly with decreasing altitude due to photochemical loss processes. Condensation is responsible for the sharp drop off in altitude of the $NH_3$ and $N_2H_4$ profiles in the few-hundred mbar range and the more localized "bite-outs" in the HCN and $CH_3CN$ profiles between $\sim$100-200 mbar.

Because of the large amount of atomic H produced from $NH_3$ photolysis in and above the ammonia condensation region, the dominant scheme for producing HCN in our Jovian photochemical model is

$$NH_3 \xrightarrow{h\nu} NH_2 + H$$

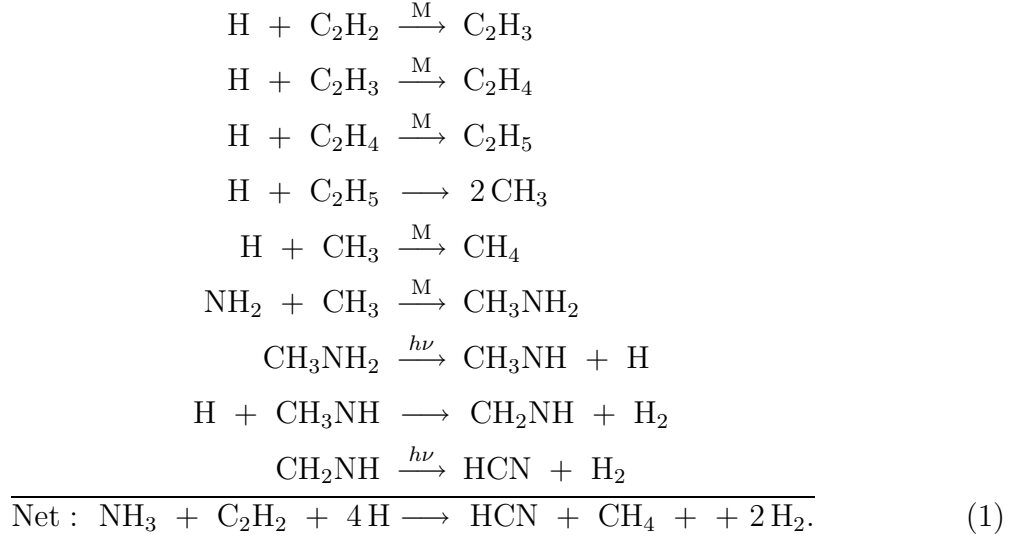

$$\text{Net}: \ NH_3 \ + \ C_2H_2 \ + \ 4\,H \longrightarrow HCN \ + \ CH_4 \ + \ + \ 2\,H_2. \tag{1}$$

The column-integrated rate of the $NH_2 + CH_3 + M \longrightarrow CH_3NH_2 + M$ reaction is about an order of magnitude larger than the next most important reaction for producing carbon-nitrogen bonds, that of $NH_2 + C_2H_3 + M \longrightarrow C_2H_5N + M$. This solution differs from theoretical models of Kaye and Strobel[35] for which the $NH_2 + C_2H_3 \rightarrow C_2H_5N$ reaction dominates, despite the fact that we use a slightly higher estimate for the rate coefficient for this reaction. The main difference between our model and that of Kaye and Strobel[35] in this regard appears to be the significance of $CH_3$ production through sequential addition of atomic H to $C_2H_x$ hydrocarbons (*i.e.*, the first half of the scheme (1) above). The sheer amount of H produced from $NH_3$ photolysis (and from $C_2H_2$ photolysis at higher altitudes) makes these three-body addition reactions for atomic H effective (see also [135]). Our solution here also differs from the results of our box-model simulations described in Section 2, for which the $NH_2 + C_2H_2 + M \rightarrow C_2H_4N + M$ reaction dominates the formation of carbon-nitrogen bonds, due to the large concentration of $C_2H_2$ and relatively high pressure in the reaction cell.

None of the carbon-nitrogen species appear to be produced in our Jovian photochemical model in large enough quantities to be observable with current technologies. After HCN, the second and third most abundant carbon-nitrogen species produced in our Jovian photochemical model are $C_2H_5NH_2$ and $CH_3CN$. The dominant mechanism for producing $CH_3CN$ in our model is

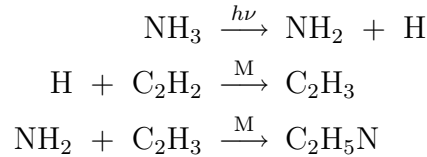

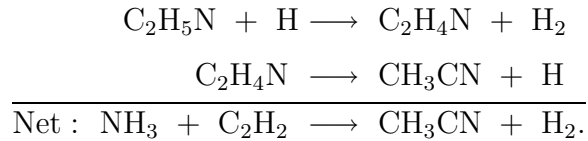

$$\text{Net}: \ NH_3 \ + \ C_2H_2 \ \longrightarrow \ CH_3CN \ + \ H_2.$$

Note that this mechanism was proposed as a way to form acetonitrile in coupled $NH_3$-$C_2H_2$ photolysis experiments.[36−38]

The dominant mechanism for producing $C_2H_5NH_2$ in our model is

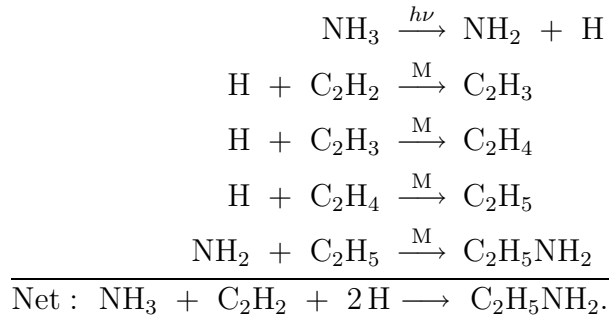

$$\text{Net}: \ NH_3 \ + \ C_2H_2 \ + \ 2\,H \ \longrightarrow \ C_2H_5NH_2.$$

This mechanism was proposed as being important for the production of ethylamine in coupled $NH_3$-$C_2H_2$ photolysis experiments.[38]

Figure 3 shows the sensitivity of some of the species abundances to the tropospheric eddy diffusion coefficient and to the assumed haze optical depth in the ammonia condensation region. A larger tropospheric eddy diffusion coefficient allows slightly more $NH_3$ to be carried up into the stratosphere, where slightly more $NH_2$ is formed as a result. The stratospheric abundances of species that depend on $NH_2$ for their production, which includes all the carbon-nitrogen species, are then slightly increased for the case of the larger $K_{zz}$, and stratospheric mole fractions are correspondingly increased. However, the larger tropospheric $K_{zz}$ also allows these species to diffuse more quickly through the bottom boundary of the model, so that the tropospheric mole fractions of species that are produced from the stratospheric coupled $C_2H_2$-$NH_3$ photochemistry are reduced when $K_{zz}$ is increased. The $NH_3$ and $N_2H_4$ abundances themselves are controlled by condensation and evaporation in the troposphere and show little sensitivity to tropospheric $K_{zz}$ values. Molecular nitrogen, on the other hand, does not condense and is produced largely in the troposphere, so increasing the tropospheric $K_{zz}$ leads to increased transport and loss through the lower boundary, resulting in a reduced $N_2$ column abundance when tropospheric $K_{zz}$ values are increased.

The dotted lines in Fig. 3 represent a model in which the eddy diffusion coefficient profile is the same as that of our nominal model (solid lines), but in which the haze vertical optical depth at 150-230 nm in the 300-700 mbar region is 8 instead of 2.7. The larger optical depth leads to increased shielding of the $NH_3$, with a corresponding slight reduction in the $NH_2$ production rate at the base of the $NH_3$ cloud and a very slight increase in the $NH_3$ mole fraction at the top of the cloud. This change in optical depth in the 300-700 mbar pressure region has almost no effect on species abundances, except for that of $N_2$, whose production rate slightly drops. Similarly, changing the assumed $NH_3$ mole fraction at the lower boundary (not shown in the figure) has almost no effect on the species abundances other than on $NH_3$ itself below its condensation region. Note that we have not included chemical loss processes in the model such as reaction of $NH_3$ with $H_2S$ or interaction with the $NH_4SH$ cloud that would explain the "stair step" reduction behavior of the $NH_3$ mole fraction in the 1-2 bar region implied by some of the observations.[107,109]

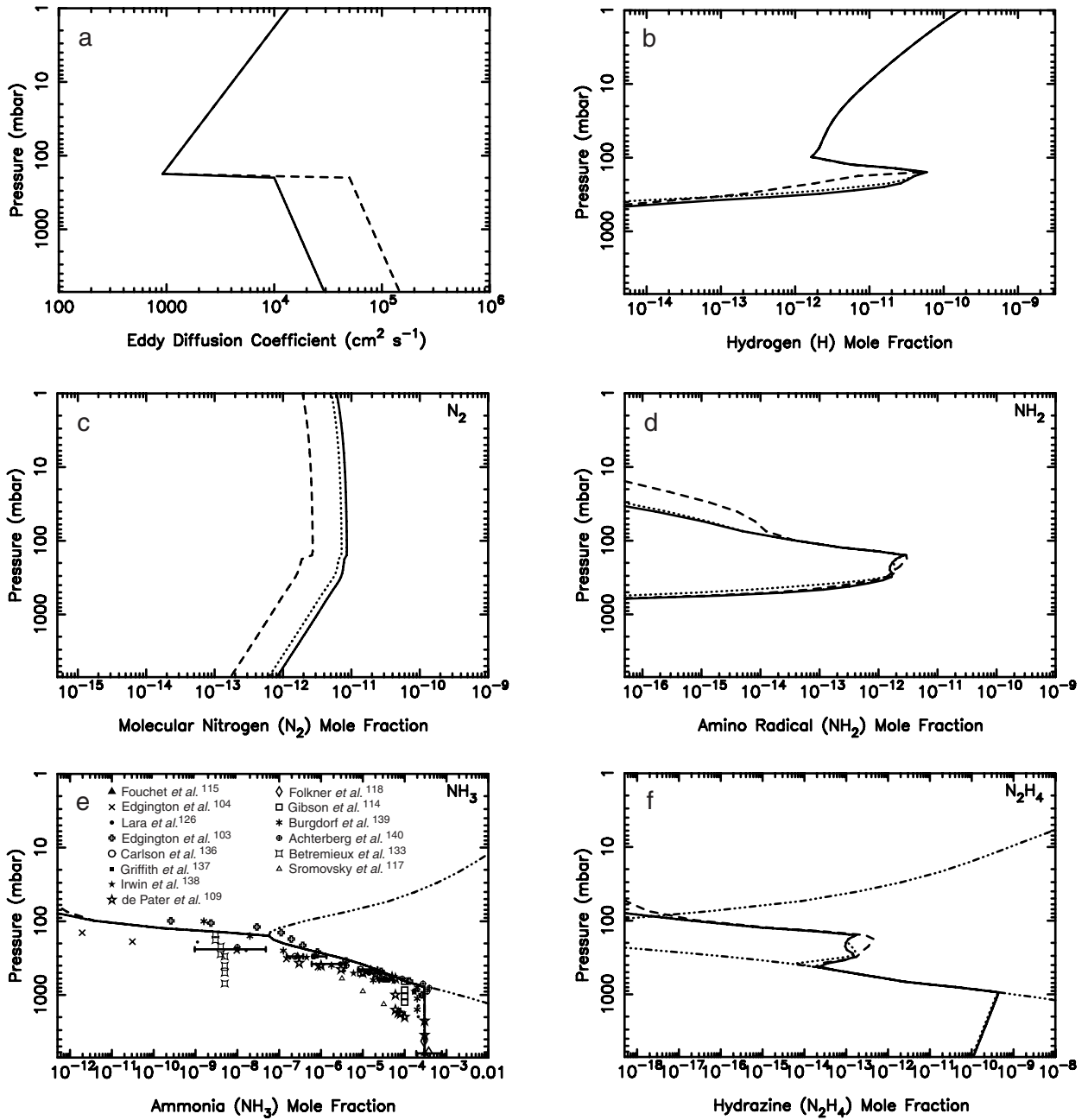

FIGURE 3. The eddy diffusion coefficient profiles assumed in the photochemical model (a), along with the mole-fraction profiles for atomic hydrogen (b), molecular nitrogen (c), amino radicals (d), ammonia (e), hydrazine (f), hydrogen cyanide (g), acetonitrile (h), methanimine (i), and ethylamine (j) in our Jovian photochemical model. The solid lines represent our nominal model, with a tropospheric eddy diffusion coefficient similar to that derived by Edgington *et al.*[104] and with a haze vertical optical depth at 150-230 nm of 2.7 in the 300-700 mbar region. The dashed lines represent a model that is the same as our nominal model, except the tropospheric $K_{zz}$ profile has been multiplied by a factor of 5. The dotted lines represent a model that is the same as our nominal model, except the haze vertical optical depth at 150-230 nm in the 300-700 mbar region is 8 instead of 2.7. The dash-triple-dot lines represent the saturation vapor density curves for the molecules in question. The model results for NH$_3$ are compared with various observations[103,104,114–118,126,136–140] in (e), and the HCN profiles are compared with the stratospheric[13] and tropospheric[15] upper limits.

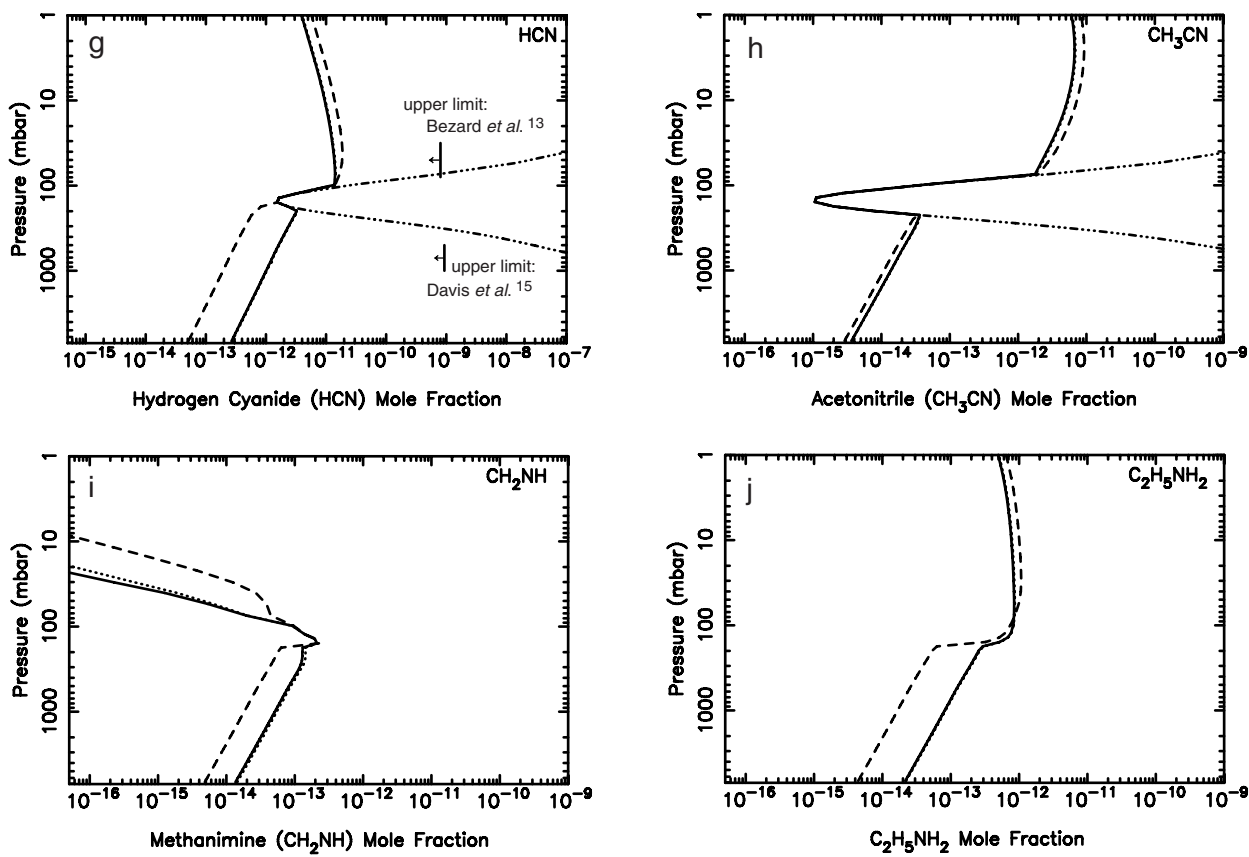

**FIGURE 3.** *(cont.)*

These simple sensitivity studies show that our main conclusion about the unimportance of coupled $C_2H_2$-$NH_3$ photochemistry on Jupiter cannot be changed by tweak-ing uncertain free parameters in the model. The observed $PH_3$ profile on Jupiter puts limits on how small the tropospheric eddy diffusion coefficient can be,[103,104,126,128] and our tropospheric $K_{zz}$ is unlikely to be much smaller than we have adopted in our nominal model. The production rate of nitriles and other organo-nitrogen compounds is mostly unaffected by the assumed optical depth within the ammonia cloud layer or the assumed $NH_3$ abundance below the cloud. Observations show $NH_3$ to be subsaturated on Jupiter, particularly in belt regions,[107] whereas our ammonia profile closely follows its saturation vapor pressure curve in its condensation region (in contrast to $N_2H_4$, which is produced rapidly enough in its condensation region that condensation loss cannot keep pace, and supersaturations are maintained). A subsaturated ammonia profile would lead to even less $C_2H_2$-$NH_3$ photochemical coupling and lower abundances of HCN and other carbon-nitrogen compounds.

The lack of coupled $C_2H_2$-$NH_3$ photochemistry in our model is entirely due to the low derived abundance of $C_2H_2$ in the $NH_3$ photolysis region in Jupiter's upper troposphere and lower stratosphere. Kaye and Strobel[35] assumed a much greater $C_2H_2$ abundance in their model — their Model A contains $C_2H_2$ concentrations a factor of $\sim$70 higher than our nominal model at their 50-km altitude upper boundary and more than four orders of magnitude greater than our nominal model throughout the troposphere. That difference is the main cause of our different predictions concerning the HCN abundance on Jupiter. On the other hand, we confirm the models of Kaye and Strobel[35] and Strobel[100] in that coupled $CH_4$-$NH_3$ photochemistry is greatly inhibited due to the physical separation of the $CH_4$ photolysis region in the upper stratosphere from the $NH_3$ photolysis region in the troposphere. The only way we could increase the net production rate of HCN in our model would be to invoke an upper tropospheric source of $C_2H_2$[24,25,132,133] or unusual dynamical conditions that allow rapid transport of stratospheric $C_2H_2$ into the troposphere; we discuss these possibilities further in Section 5.2.

## 4. Thermochemical Kinetics and Transport Model for the Deep Troposphere

At very high temperatures ($\gtrsim$1500 K) in Jupiter's deep troposphere, the atmospheric composition is controlled by thermochemical equilibrium. Equilibrium models that include nitrogen species[141−143,19−21] show that $NH_3$ is the dominant nitrogen-bearing constituent throughout the Jovian atmosphere; HCN and $N_2$ are not very abundant at colder high altitudes, but the equilibrium mole fractions of $N_2$ and HCN are both expected to increase toward the deeper, hotter regions of the troposphere. For the observable regions in the upper troposphere, the equilibrium HCN abundance, in particular, is negligible. However, as was first discussed quantitatively by Prinn and Barshay,[144] thermochemical equilibrium may be difficult to maintain on Jupiter in the presence of rapid convective mixing. In this scenario, thermochemical equilibrium can be preserved only as long as the chemical kinetic time scale for conversion between different molecular species is shorter than the time scale for vertical atmospheric mixing. As a parcel of gas from deeper, hotter levels is transported up to cooler atmospheric regions, it will eventually encounter regions where the chemical kinetic conversion time scale becomes longer than the transport time scale, at which point the mole fractions of species like CO, $N_2$, $CH_3NH_2$, or HCN can become "quenched" or "frozen in" due to the inability of the kinetic reactions to overcome activation energy barriers. At altitudes above that quenching level (*i.e.*, at altitudes above the level where the chemical-kinetic time constant equals the convective mixing time constant), the mole fraction of the quenched species will remain fixed at the equilibrium abundances achieved at the quench point.[144] Because of this transport-induced disequilibrium process, species can be present in the upper troposphere of Jupiter in abundances much greater than their predicted equilibrium abundances. The disequilibrium quenching of $N_2$ and HCN on Jupiter has been discussed by several investigators[145,10,20,21] (see also the relevant discussions for other solar-system applications[146−148]).

Through time-constant and quenching-level arguments, Lewis and Fegley[10] suggest that HCN would quench at $\sim$1200 K on Jupiter, resulting in a quenched steady-state mole fraction for HCN of only $\sim$1 $\times$ 10$^{-12}$ for a solar-composition gas. Lewis and Fegley,[10] using arguments from Prinn and Fegley,[147] suggest that the rate-limiting step responsible for quenching the HCN abundance is the reaction $H_2 + HCN \leftrightarrow CH_2 + NH$. The rate coefficient for this reaction and its reverse have never been measured; Lewis and Fegley[10] assume that the reverse reaction will proceed rapidly with a rate coefficient of $1 \times 10^{-10}$ cm$^3$ s$^{-1}$. The rate coefficient in the forward direction is then estimated from the equilibrium constant of the reaction, along with the reverse rate coefficient, using the principle of microscopic reversibility. Fegley and Lodders[20] make these same assumptions in their follow-up study, but they derive a much larger quenched HCN mole fraction of 0.6-2.6 ppb for Jupiter. The differences between the two results are attributed to the larger elemental enrichment factors used in the more recent study[20,21]. We note that the most likely products of the $CH_2$ + NH reaction could be $H_2CN$ + H and/or N + $CH_3$, rather than HCN + $H_2$, so that the overall $H_2$ + HCN reaction as stated might be an oversimplification of a multiple-step process, but otherwise the assumptions seem reasonable. Is that reaction the only available mechanism for HCN loss, though? That premise seems unlikely. The reaction between HCN and $H_2$ is very endothermic and will be exceedingly slow, even at 1500 K, and we suspect there are more effective HCN destruction mechanisms under deep-tropospheric conditions on Jupiter. Moreover, as is demonstrated by Smith,[46] the adoption of the pressure scale height $H$ for the characteristic length scale in the expression for the convective mixing time constant leads to a roughly two-order-of-magnitude overestimation of the transport time scale used in the time-constant approach[144,10,20,21] to predict quenched disequilibrium abundances on Jupiter. If Smith[46] is correct, the HCN quenching level is likely even deeper in the atmosphere than Fegley and Lodders[20] have assumed (given their rate-limiting mechanism), which in turn suggests an even higher predicted quenched HCN mole fraction — a result in clear violation of the HCN upper limits.[13-15]

The details of the original Prinn and Barshay[144] time-constant approach for CO quenching on Jupiter have been questioned.[24,45-49] Both the assumed rate-limiting step (and its estimated rate coefficient) and the assumptions regarding the transport time scale have been criticized; several groups have suggested ways in which the original assumptions could be improved.[24,45-49]. Fueled by these criticisms and suggested improvements, we have recently developed a way to bypass the back-of-the-envelope time-scale approach by directly modeling chemical kinetics and transport in the deep troposphere of Jupiter to more quantitatively investigate carbon-hydrogen-oxygen chemistry and the transport-induced quenching of disequilibrium C-H-O species. The results are presented in Visscher *et al.*[49] Using this model, we confirm the results of Smith[46] regarding the transport time constants and identify the most likely rate-limiting step for the quenching of CO on Jupiter. The new rate-limiting mechanism suggested by Visscher *et al.*[49] helps resolve a long-standing controversy regarding the Prinn and Barshay[144] scheme and the origin of tropospheric CO on Jupiter, as well as helps constrain the deep water abundance on Jupiter. Based on kinetics and diffusion only, the model is able to reproduce the equilibrium composition at deep, hot atmospheric levels, and then transitions smoothly to a quenched regime at higher altitude levels based on the rates of the reactions controlling the interconversion of the different atmospheric constituents.

For this paper, we discuss the results regarding nitrogen chemistry in the deep Jovian troposphere.

## 4.1. Thermochemical model description

We again use the Caltech/JPL KINETICS code[43] to solve the continuity equations for the atmospheric constituents, but we focus this time on Jupiter's deep troposphere. The model extends from 12,650 bar (2500 K) to 17.4 bar (399 K) in a grid of 144 atmospheric levels, with a vertical resolution of at least twenty altitude levels per scale height. The assumed pressure-temperature profile in the 17-24 bar region is taken from the Galileo probe ASI data[69] and is extended to greater depths along an adiabat, assuming ideal-gas behavior.[148] Thermochemical equilibrium is adopted as an initial condition, with elemental abundances taken from the Galileo Probe Mass Spectrometer (GPMS) results[110,112] for carbon, nitrogen, and helium, but not for oxygen. The oxygen elemental abundance determined by the GPMS is considered to be a lower limit to the deep Jovian abundance due to the probe's entry into an anomalous "hot-spot" region.[107] The oxygen abundance in our model is set at 2.6 times the assumed protosolar value of $H_2O/H_2 = 9.61 \times 10^{-4}$,[149] where it is assumed here that a portion of the total oxygen content has already been removed by rock-forming elements (which we do not consider in the model). This level of oxygen enrichment was found by Visscher $et\ al.$[49] to provide a good fit to the observed tropospheric CO mole fraction.[48] The NASA CEA code[150] is used to calculate thermodynamic equilibrium, with thermodynamic parameters taken from Gurvich $et\ al.$,[151] Chase,[152] Burcat and Ruscic,[153] and other literature sources. Zero flux boundary conditions are adopted at the top and bottom of the model such that no mass enters or leaves the system. Transport is assumed to occur through vertical eddy diffusion, with our nominal model adopting a constant tropospheric $K_{zz}$ of $1 \times 10^8$ cm$^2$ s$^{-1}$ (see Visscher $et\ al.$[49] for further details).

We use a subset of ~1800 reactions and ~120 species from our photochemical model described in Sections 2 and 3. Because we must fully reverse all our reactions (using the principle of microscopic reversibility) in order to accurately reproduce equilibrium compositions with the kinetic model, we are forced to omit several of the complex organo-nitrogen compounds observed in coupled $C_2H_2$-$NH_3$ photolysis experiments[36−39] due to a lack of information on thermodynamic properties (see Section 2 for a list of these species). The top of our model (17 bar) is also deep enough that ultraviolet photons do not penetrate, so we neglect photolysis reactions. Other than these changes, the rate coefficient expressions are taken from the photochemical model. The kinetics of nitrogen species under deep-tropospheric conditions on Jupiter is less well understood than the corresponding case for C-H-O species, and our predicted quench levels for the nitrogen-bearing species will be correspondingly less certain. We deem the modeling exercise worth the effort, however, if for nothing else than to suggest reactions that could be important for the dominant quenching mechanisms and thus warrant further study. More importantly, we hope to resolve the apparent paradox mentioned above regarding the large quenched disequilibrium abundance of HCN in excess of observational limits (*cf.*[15,20,21]) that is predicted when the Lewis and Fegley[10] rate-limiting kinetic mechanism for HCN destruction is used in combination with the convective-mixing length scale derived by Smith.[46]

## 4.2. Thermochemical model results

Figure 4 shows the results of our thermochemical kinetics and transport model for some of the major nitrogen-bearing species. Because our kinetic model considers the same species as our thermochemical-equilibrium calculations and because we fully reverse all of our kinetic reactions using that same thermodynamic data, our kinetic model will reproduce the equilibrium results in the absence of transport, provided that we allow enough time for equilibrium to be achieved at any particular temperature. At hot, deep levels in the kinetics-transport model shown in Fig. 4, the kinetic reactions are very fast, and equilibrium is maintained. However, the species profiles diverge from equilibrium (shown with dashed lines in Fig. 4) when the kinetic rates can no longer keep up with dynamical mixing (*i.e.*, when the transport time scale falls below the kinetic loss time scale for the species in question). This divergence occurs in the cooler, upper regions of the model: many reactions have significant activation barriers that cannot readily be overcome as temperatures drop, so that the reactions proceed predominantly in one direction, and equilibrium is not preserved. At altitudes above this quench level, the mole fraction of a quenched species remains constant. This behavior was first predicted analytically by Prinn and Barshay.[144] Note that different species shown in Fig. 4 have different quench levels. Molecular nitrogen, with its strong triple bond, is difficult to destroy kinetically and so is quenched at relatively hot, deep levels, whereas HCN equilibrium is preserved even at relatively cool temperatures such that the quench level is at higher altitudes. Some species, such as HNCO and $CH_2NH$, exhibit more complex behavior due to the effects of the quenching of other constituents. For instance, HNCO first begins to diverge from equilibrium when CO quenches, but HNCO itself does not quench until it reaches higher altitudes. In the intervening regions, HNCO continues to maintain an equilibrium with the quenched CO. Similarly, CH$_2$NH diverges from equilibrium when HCN quenches but does not itself fully quench until it reaches higher altitudes.

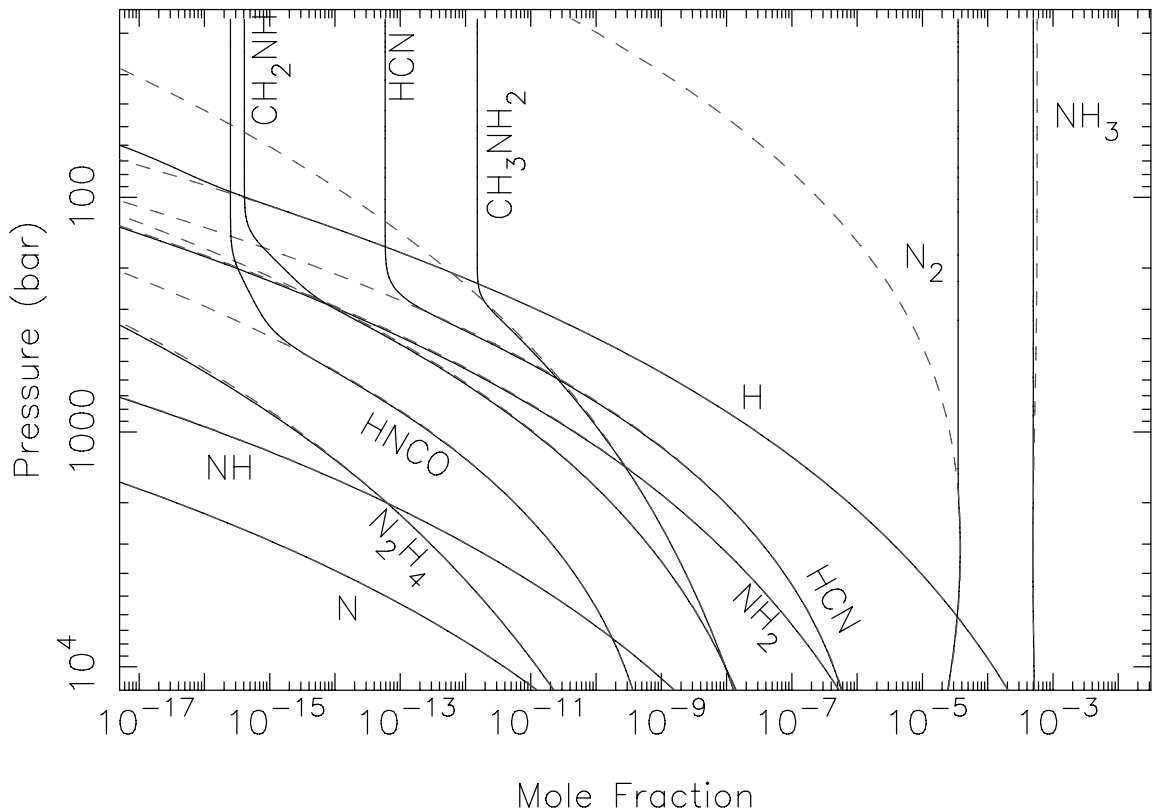

FIGURE 4. Mole-fraction profiles (as labeled) for several nitrogen-bearing species in our Jovian deep-troposphere thermochemical kinetics and transport model for an assumed constant eddy diffusion coefficient of $1 \times 10^8$ cm$^2$ s$^{-1}$. The dashed lines show our thermochemical equilibrium solution. Note that kinetic reactions are so fast in the hotter, deeper regions of the model that equilibrium can be maintained. However, the species abundances diverge from equilibrium and are quenched at colder, higher levels as the transport time scale drops below the kinetic time constants for conversion between the different species.

Based on our adopted nitrogen reaction mechanism, we find that HCN does not quench until it reaches the $\sim$880-K, 260-bar level. As a result, the quenched mole fraction is only $\sim$6 $\times 10^{-14}$, a value well below the observational limit for HCN of 0.93 ppb in the Jovian upper troposphere.[15]  The dominant mechanism for HCN loss in our model is the following scheme:

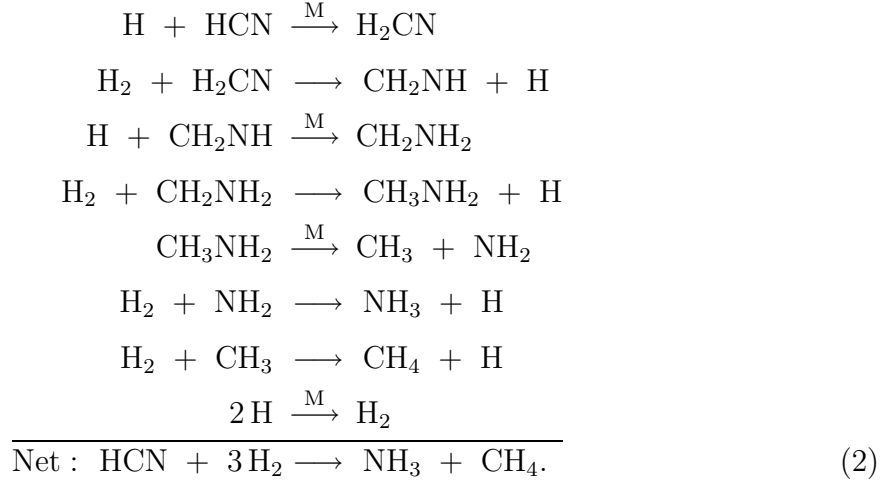

$$
\begin{aligned}
\mathrm{H} + \mathrm{HCN} &\xrightarrow{\mathrm{M}} \mathrm{H_2CN} \\
\mathrm{H_2} + \mathrm{H_2CN} &\longrightarrow \mathrm{CH_2NH} + \mathrm{H} \\
\mathrm{H} + \mathrm{CH_2NH} &\xrightarrow{\mathrm{M}} \mathrm{CH_2NH_2} \\
\mathrm{H_2} + \mathrm{CH_2NH_2} &\longrightarrow \mathrm{CH_3NH_2} + \mathrm{H} \\
\mathrm{CH_3NH_2} &\xrightarrow{\mathrm{M}} \mathrm{CH_3} + \mathrm{NH_2} \\
\mathrm{H_2} + \mathrm{NH_2} &\longrightarrow \mathrm{NH_3} + \mathrm{H} \\
\mathrm{H_2} + \mathrm{CH_3} &\longrightarrow \mathrm{CH_4} + \mathrm{H} \\
\mathrm{2\,H} &\xrightarrow{\mathrm{M}} \mathrm{H_2} \\
\hline
\mathrm{Net:\ HCN} + \mathrm{3\,H_2} &\longrightarrow \mathrm{NH_3} + \mathrm{CH_4}.
\end{aligned}
\tag{2}
$$

The rate-limiting step in this scheme is the $\mathrm{H_2} + \mathrm{H_2CN} \rightarrow \mathrm{CH_2NH} + \mathrm{H}$ reaction. Our rate coefficient for this reaction comes from the reverse reaction, whose rate coefficient has been calculated from the DHT method.[60]  Although the rate coefficient for the $\mathrm{H_2} + \mathrm{H_2CN} \rightarrow \mathrm{CH_2NH} + \mathrm{H}$ reaction is a relatively low $\sim 3 \times 10^{-18}$ cm$^3$ s$^{-1}$ at the quench level, it is still much faster than the likely rate for the $\mathrm{H_2} + \mathrm{HCN}$ reaction suggested as the rate-limiting step by Lewis and Fegley[10] and subsequent modelers[20,21] because of the relative reactivity of $\mathrm{H_2CN}$ versus HCN. Note that our scheme begins with the three-body addition of H to HCN, and continues with the products reacting with H and/or $\mathrm{H_2}$, which are abundant in the Jovian troposphere, to form a hydrogen-saturated single-bonded species, which can then thermally dissociate to break the C-N bond. A very similar scheme was invoked by Visscher *et al.*[49] to explain the dominant mechanism for breaking the strong carbon-oxygen bond and destroying CO in the Jovian deep troposphere (see also[45]), and we find $\mathrm{N_2}$ to be destroyed by a similar process.

For instance, the dominant scheme leading to $\mathrm{N_2}$ destruction in our model is

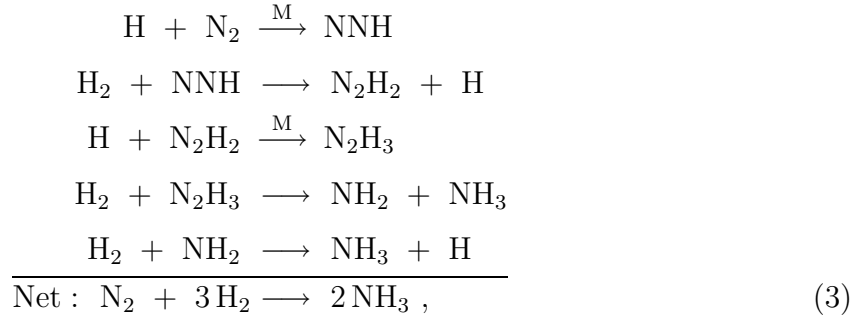

$$
\begin{aligned}
\mathrm{H} + \mathrm{N_2} &\xrightarrow{\mathrm{M}} \mathrm{NNH} \\
\mathrm{H_2} + \mathrm{NNH} &\longrightarrow \mathrm{N_2H_2} + \mathrm{H} \\
\mathrm{H} + \mathrm{N_2H_2} &\xrightarrow{\mathrm{M}} \mathrm{N_2H_3} \\
\mathrm{H_2} + \mathrm{N_2H_3} &\longrightarrow \mathrm{NH_2} + \mathrm{NH_3} \\
\mathrm{H_2} + \mathrm{NH_2} &\longrightarrow \mathrm{NH_3} + \mathrm{H} \\
\hline
\mathrm{Net:\ N_2} + \mathrm{3\,H_2} &\longrightarrow \mathrm{2\,NH_3} ,
\end{aligned}
\tag{3}
$$

where NNH has recently been recognized as an important intermediate in the combustion chemistry of nitrogen species, particularly in flame fronts where the concentration of atoms is high.[60] Note the similarity of the first three steps in this reaction scheme (3) to the first three steps in our dominant HCN destruction scheme (2) above. The rate-limiting step in this $N_2$ destruction scheme (3) is $H + N_2H_2 + M \rightarrow N_2H_3 + M$. Our rate coefficient for this reaction again comes from the reverse reaction, whose rate coefficient was derived from QRRK analysis by Dean and Bozzelli.[60]

As with the destruction of other species with strong bonds in our model, our proposed mechanism begins with H addition to the very stable $N_2$ molecule, followed by sequential reactions of $H_2$ and H to form a single-bonded N-N species, before the N-N bond is broken. This mechanism differs significantly from the $N_2 + H_2 \rightarrow 2\,NH$ gas-phase, rate-limiting mechanism suggested by previous investigators.[145−147,10,20,21] Our mechanism leads to somewhat more effective conversion of $N_2 \rightarrow NH_3$ in Jupiter's deep troposphere; however, given the near-vertical slope of the $N_2$ equilibrium profile at depth, our prediction for the quenched disequilibrium $N_2$ mole fraction of $3.5 \times 10^{-5}$ does not differ too much from the $2\text{-}3 \times 10^{-5}$ mole fraction predicted by Fegley and Lodders,[20] who assumed a slightly smaller nitrogen elemental abundance in their model (see also[21]). In fact, the assumptions about the deep nitrogen elemental abundances on Jupiter have the largest effect on the predictions concerning the mole fraction of $N_2$ dredged up from the deep atmosphere.[20] No matter what the actual rate-limiting step is, we agree with the conclusions of Prinn and Olaguer[145] and subsequent modelers that $N_2$ is likely to be the most abundant quenched disequilibrium species in the upper troposphere of Jupiter.

Note that the conversion of $N_2$ to $NH_3$ might also occur heterogeneously on the surface of metallic iron grains,[145] as in the industrial Haber process. If catalytic $N_2$ destruction on grain surfaces is occurring on Jupiter, Prinn and Olaguer[145] demonstrate that this process could be more efficient than the pure gas-phase mechanism we consider, leading to a reduced, but still significant, $N_2$ mole fraction of 0.3-6 ppm in their model. However, modern chemical equilibrium calculations[20,21,154] demonstrate that Fe is removed from the atmosphere by condensation at altitudes much deeper than the level where the catalysis would be occurring on Jupiter, making metallic iron unavailable as a catalyst for $N_2 \rightarrow NH_3$ conversion. The presence of another suitable catalyst remains problematic, and homogeneous gas-phase reactions are expected to dominate.[20]

Molecular nitrogen is stable in the Jovian upper troposphere: $N_2$ does not readily react with the photolytic products of $NH_3$ and hydrocarbon photochemistry, and it is shielded from photolysis by the large overlying column of $H_2$. We expect $N_2$ to survive well into the upper stratosphere, where its mole fraction will eventually be reduced due to molecular diffusion. Galactic cosmic rays may initiate interesting $N_2$ chemistry, as suggested for Neptune,[155] and $N_2$ may participate in interesting ionospheric chemistry, as on Titan.[156]

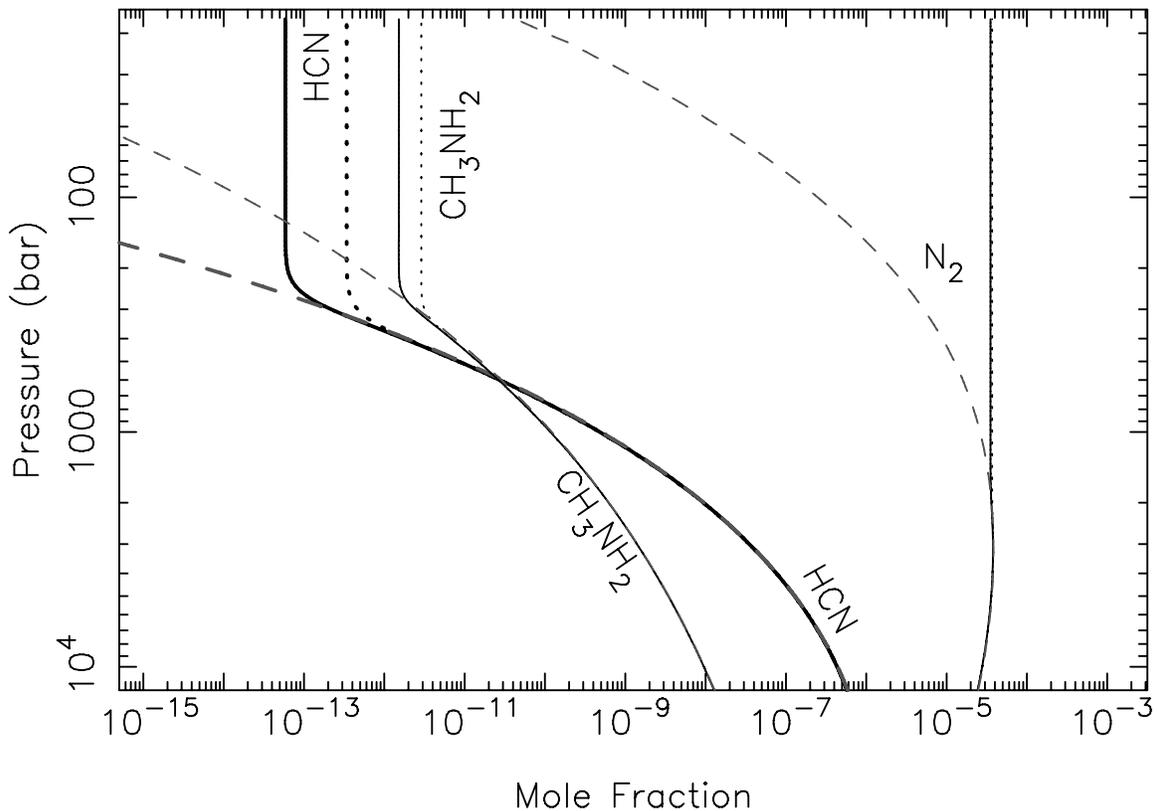

FIGURE 5. Mole-fraction profiles for $N_2$, HCN, and $CH_3NH_2$ in our nominal thermochemical kinetics and diffusion model with $K_{zz} = 1.0 \times 10^8$ cm$^2$ s$^{-1}$ (solid lines) compared with a model with $K_{zz} = 1.0 \times 10^9$ cm$^2$ s$^{-1}$ (dotted lines). The dashed lines show the thermochemical equilibrium solution.

As originally noted by Lewis and Fegley,[10] the quenched mole fraction of methylamine ($CH_3NH_2$) is greater than that of HCN, so that the additional deep tropospheric source of $CH_3NH_2$ could enhance photochemical production of HCN (see the last part of scheme (1) above). However, the upper-tropospheric quenched mole fraction of $CH_3NH_2$ in our model is only $1.5 \times 10^{-12}$. Even if the conversion of $CH_3NH_2$ into HCN were 100% effective, the HCN produced from $CH_3NH_2$ photochemistry would still be well below the HCN upper limits.[13−15]

Fegley and Lodders,[20] Lodders and Fegley,[21] and Visscher *et al.*[49] demonstrate that the mole fractions of the quenched species on Jupiter are very sensitive to the adopted eddy diffusion coefficient in the vicinity of the quench level. Our adopted nominal $K_{zz}$ value of $1 \times 10^8$ cm$^2$ s$^{-1}$ is based on free-convection and mixing-length theory for a rapidly rotating atmosphere.[157,158,49] The exact value of the eddy diffusion coefficient is uncertain by about an order of magnitude. A larger value would lead to more rapid mixing and a quench level deeper in the atmosphere, with correspondingly higher values for the mole fractions of quenched disequilibrium species like HCN, $CH_3NH_2$, and CO. Figure 5 illustrates how the quenched abundances of $N_2$, HCN, and $CH_3NH_2$ would change for a larger assumed $K_{zz}$ of $1 \times 10^9$ cm$^2$ s$^{-1}$. Even with this maximum value of $K_{zz}$, the predicted upper-tropospheric mole fraction for HCN is well below the observational upper limit. Note that the equilibrium gradient in the deep troposphere controls how the quenched abundance will be affected by the eddy diffusion coefficient; $N_2$, with its nearly vertical profile near the quench levels, is relatively unaffected, whereas HCN is strongly affected.

One loss mechanism for HCN that is not in the model is reaction of HCN with $NH_3$ to form condensed $NH_4CN$ salt, as was originally suggested by Lewis.[143] The formation of this species may significantly limit the possible vapor abundance of HCN at temperatures below $\sim$160 K (*i.e.*, above 0.9 bar) on Jupiter.

## 5. Discussion

Our modeling suggests that neither the photochemical source nor the deep tropospheric source can provide much HCN to the Jovian upper troposphere — a result that is consistent with the low observational upper limit for non-cometary HCN on Jupiter.[13−15] Our conclusions are at odds with some of the previous modeling predictions found in the literature,[35,20,21] and we now discuss some of the reasons for the differences.

### 5.1. Deep tropospheric HCN source compared with previous models

The quenched-disequilibrium model that can be most directly compared with ours is that of Fegley and Lodders,[20] due to the greater-than-solar nitrogen elemental abundance assumed in both models: we assume a $NH_3/H_2$ ratio of $6.64 \times 10^{-4}$,[112] whereas Fegley and Lodders[20] assume a $NH_3/H_2$ ratio of $5.2 \times 10^{-4}$. Fegley and Lodders[20] use a time-constant approach rather than thermochemical kinetics and transport models, but we can use similar arguments for our comparisons. Based on an earlier estimate from Prinn and Fegley[147], Fegley and Lodders[20] assume that the rate-limiting step in the reduction of HCN is the reaction HCN + H$_2$ → CH$_2$ + NH, with an estimated rate coefficient of $1.08 \times 10^{-8} \exp(-70,456/T)$ cm$^3$ s$^{-1}$, for $T$ in K. The estimated chemical kinetic time constant for HCN destruction is then

$$\tau_{chem} = \left(1.08 \times 10^{-8} \exp(-70,456/T)\,[\text{H}_2]\right)^{-1} \; .$$

If one then assumes that the convective mixing time scale $\tau_{mix} = L^2/K_{zz}$, with the effective convective length scale $L$ being the atmospheric pressure scale height $H$, then $\tau_{mix}$ is equal to $\tau_{chem}$ at level at which the equilibrium HCN mole fraction is ~$1 \times 10^{-9}$ in the Fegley and Lodders model[20] — a value similar to the observed HCN upper limit of 0.93 ppb.[15] However, Smith[46] argues that the effective mixing length scale $L$ is more like $0.11H$, such that $\tau_{mix}$ would be equal to $\tau_{chem}$ deeper in the atmosphere, near the 2300-bar, 1630-K level in our model, at which point the equilibrium HCN mole fraction would be ~$1.6 \times 10^{-8}$. This value is almost twenty times the observational upper limit for HCN and is clearly not supported by the infrared and submillimeter observations.[13-15] Thus, we have an apparent paradox in that the predicted HCN mole fraction from the quenched deep-tropospheric source can only be consistent with the observational upper limits if one ignores the modeling of Smith.[46]

We propose a resolution to this apparent paradox. Our model contains several more efficient pathways to HCN → NH$_3$ conversion than the HCN + H$_2$ → CH$_2$ + NH rate-limiting step first proposed by Prinn and Fegley.[147] Scheme (2) above shows the most effective HCN → NH$_3$ conversion scheme in our model, but there are also several others that are more efficient than the Prinn and Fegley mechanism.[147] The rate-limiting step in scheme (2) is the reaction

$$\text{H}_2 + \text{H}_2\text{CN} \rightarrow \text{CH}_2\text{NH} + \text{H} \; , \qquad (4)$$

for which the rate coefficient in our model is determined from the reverse reaction.[60] We can fit an Arrhennius expression to our rate coefficient for reaction (4), as calculated from the reverse reaction: $k_4 = 1.011 \times 10^{-18}\,T^{1.941} \exp(-10,682.5/T)$ cm$^3$ s$^{-1}$, for $T$ in K. If we use this reaction as our rate-limiting step, the kinetic time constant for HCN destruction is

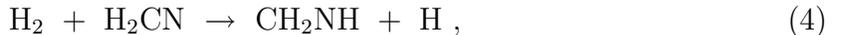

$$\tau_{chem} = \frac{[\text{HCN}]}{d[\text{HCN}]/dt} = \frac{[\text{HCN}]}{k_4\,[\text{H}_2]\,[\text{H}_2\text{CN}]} \; .$$

Then, using equilibrium calculations along with an effective length scale of $L \approx 0.11H$,[46] we find that $\tau_{chem}$ equals $\tau_{mix}$ at the $\sim$870-K, $\sim$250-bar level in our model, where the equilibrium HCN mole fraction is $4 \times 10^{-14}$ from this time-constant approach, compared to our model-derived value of $6 \times 10^{-14}$. The similarity in the results from the time-constant approach using the Smith[46] length scale as compared with our thermochemical kinetics and transport model results suggests that (a) the Smith[46] results are indeed reasonable and should not be ignored (as they frequently are now) by those who use time-constant arguments to calculate the abundance of quenched disequilibrium species dredged up from the deep troposphere, and (b) the time-constant approach is valid (*i.e.,* one does not need a full thermochemical kinetics and transport model to predict quenched disequilibrium abundances on the giant planets). Visscher *et al.*[49] came to similar conclusions from their modeling of CO reduction on Jupiter (see also Bézard *et al.*[48]). Furthermore, the fact that our derived quenched HCN mole fraction lies comfortably below the observational upper limits suggests that there are faster mechanisms for converting HCN to $NH_3$ in the Jovian troposphere than the $HCN + H_2 \rightarrow CH_2 + NH$ reaction originally proposed by Prinn and Fegley.[147]

Our dominant schemes for the reduction of HCN, $N_2$, and CO (see Visscher *et al.*[49]) on Jupiter all start with H-atom addition, followed by reaction of the $H_2CN$, NNH, or HCO adducts with $H_2$, and subsequent reactions with H and/or $H_2$ to eventually form species with single C-N, N-N, or C-O bonds, before those bonds are finally broken. We thus find alternative pathways to the $H_2$-plus-strongly-bonded-constituent reactions that form the backbone of the mechanisms proposed by Prinn and Barshay,[144] Prinn and Fegley,[147] Prinn and Olaguer,[145] and subsequent modelers. Yung *et al.*[45] call such reactions "kinetically too ambitious" — a wonderful phrase that has been often quoted. Our mechanism suggests that other, less ambitious reactions can do the job.

Not all the reactions in our full mechanism have firmly determined rate coefficients, and we may be missing important pathways and/or species in our reaction list. As such, we cannot be completely confident in our derived abundances for the quenched disequilibrium species. However, the model development is based on the best reaction rate coefficients available today from combustion-chemistry studies and terrestrial atmospheric chemistry studies (both of which are unfortunately concerned more with the oxidation of reduced species than the reduction of oxidized species) and are likely significant improvements over the $H_2 + N_2$ and $H_2 + HCN$ mechanisms proposed 30 years ago. Given the likely importance of such processes for extrasolar giant planets, as well as for giant planets within our own solar system, we encourage further study of the dominant mechanisms for the reduction of $N_2$ and HCN in hydrogen-dominated atmospheres.

## 5.2. Photochemical HCN source and the tropospheric $C_2H_2$ abundance

Our modeling in Section 3 demonstrates that $NH_3$-$C_2H_2$ photochemical coupling is not a significant source of HCN on Jupiter. Photochemical destruction of $C_2H_2$ in the lower stratosphere limits the amount of acetylene that diffuses into the $NH_3$ photolysis region, and both condensation and photolysis limit the amount of $NH_3$ that diffuses into the stratosphere. Ammonia and acetylene are simply not present in large enough quantities together to provide a source of HCN. In contrast, Kaye and Strobel[35] predict that as much as a few ppb HCN could form in the upper troposphere from coupled $NH_3$-$C_2H_2$ photochemistry. As we mention above, the main difference in our models relates to the $C_2H_2$ abundance rather than to major differences in the kinetic reaction-rate coefficients adopted for the first critical pathway for formation of carbon-nitrogen bonds. For their model in which a few ppb of HCN is formed, Kaye and Strobel[35] assume a fixed acetylene distribution in which $C_2H_2$ is uniformly mixed with altitude at a mixing ratio of $3 \times 10^{-8}$, based on observations available at the time,[159] whereas our model predicts $C_2H_2$ mole fractions of 1 ppb at $\sim$80 mbar, with rapidly increasing mole fractions at higher altitudes and decreasing mole fractions at lower altitudes (see Fig. 2). However, even with the older observations,[159] the mixing-ratio profiles that provided the best fit to the observed spectra required $C_2H_2$ to be depleted below the 100-mbar level, suggesting little $C_2H_2$ in the troposphere. More recent ultraviolet and infrared observations clearly indicate that the $C_2H_2$ mole fraction decreases with decreasing altitude in Jupiter's stratosphere,[13,106,115,160,161] firmly pointing to an upper stratospheric photochemical source for the acetylene. Moreover, a severe constraint on the tropospheric $C_2H_2$ mole fraction is indicated by the lack of absorption wings observed for the resolved $C_2H_2$ line profiles in thermal-infrared spectra.[133,161] Thus, the fact that all recent Jovian hydrocarbon photochemical models predict a rapidly decreasing $C_2H_2$ mole fraction in the lower stratosphere and into the troposphere[50,106,125,135,161−167] appears consistent with present-day observations.[106]

In contrast, two sets of observations have been used to suggest that the $C_2H_2$ mole fraction is relatively large in the Jovian troposphere. The first observation consists of ultraviolet spectra from the Jovian equatorial region taken with the Faint Object Spectrograph (FOS) onboard the Hubble Space Telescope (HST).[132,133] The acetylene profile derived from the Bétrémieux and Yelle[132] and Bétrémieux *et al.*[133] analyses of this data exhibit a mole fraction that decreases with decreasing altitude in the stratosphere but then increases again in the troposphere, for a best fit mole fraction of $1.5 \times 10^{-7}$ in the 120-700 mbar region. This large increase in the troposphere has been used as evidence that lightning and thundershock sources of acetylene production exist in the Jovian troposphere.[168] The second observation is the Galileo probe mass spectrometer data[110,112] that are interpreted as being consistent with ethane and other non-methane hydrocarbons going through a mole-fraction minimum near the 1-bar level, followed by an increased mole fraction at pressures greater than 16 bar.[169] Hunten[169] interprets the GPMS signature as being caused by efficient ethane and acetylene adsorption on stratospheric and tropospheric aerosols, which then rain down through the atmosphere until they reach temperatures high enough for desorption to occur.

We find these reports of large tropospheric mole fractions of $C_2H_2$ and/or $C_2H_6$ to be unconvincing, at least for the bulk of the planet. The HST/FOS ultraviolet spectra are affected by scattering and absorption from many gas-phase and aerosol species, all of which have poorly constrained parameters. Both $NH_3$ and $C_2H_2$ are clearly detected in the HST/FOS ultraviolet spectra,[132,133] but deriving mole fractions from the spectra may be problematic. The strongest argument against the $1.5 \times 10^{-7}$ tropospheric $C_2H_2$ mole fraction derived by Bétrémieux *et al.*[133] is that such a large amount would generate absorption wings around observed mid-infrared $C_2H_2$ emission lines, which Bétrémieux *et al.* recognize is inconsistent with their own ground-based infrared data,[133] as well as with other mid-infrared observations.[115,161] The $NH_3$ profile derived from the HST/FOS dataset[133] also differs considerably from profiles inferred from other observations (see Fig. 3e) for reasons that are unclear. Bétrémieux *et al.*[133] suggest that their unexpectedly large derived $C_2H_2$ tropospheric mole fraction, as well as their unexpectedly small derived $NH_3$ mole fraction, could result from uncertain $NH_3$ ultraviolet absorption cross sections at relevant Jovian temperatures, from some kind of dynamical situation where the temperature profile masks the $C_2H_2$ infrared absorption wings, or from tropospheric $NH_3$ and $C_2H_2$ profiles that vary with location and/or time on Jupiter. We note that the spatially resolved Cassini Composite Infrared Spectrometer (CIRS) data of Nixon *et al.*[161] preclude such large $C_2H_2$ mole fractions over all the latitude regions that were investigated by CIRS, although given the size of the HST/FOS footprint, the explanation of an anomalous localized atmospheric region may still be possible.

In any case, the global-average Infrared Space Observatory data of Fouchet *et*

*al.*[115] and the spatially resolved Cassini/CIRS data analyzed by Nixon *et al.*[161] demonstrate that the spectral shape of the thermal-infrared $C_2H_2$ features is inconsistent with large tropospheric $C_2H_2$ mole fractions being present over the bulk of Jupiter.

The increase in the mole fraction of heavy hydrocarbons in the GPMS data described by Hunten[169] occurs too deep to affect infrared spectra, and no remote-sensing observations currently exist to test this claim. However, the suggestion[169] that the GPMS data are consistent with the adsorption of non-methane hydrocarbons onto fluffy stratospheric and upper-tropospheric aerosols that then sediment into the troposphere to evaporate at high temperatures has been questioned by Wong,[170] based on the extensive GMPS calibrations from his thesis.[171] Wong[170] states that the GPMS data "do not support the vertical variation of ethane mixing ratio" that is key to the Hunten[1968] aerosol adsorption/desorption model. In particular, the GPMS data do not support an upper-tropospheric minimum in the ethane mole fraction, nor do they indicate that the mole fraction increases with depth in the troposphere. Wong[170] provides evidence, including the relative abundances of the measured species, that suggests that the non-methane hydrocarbons measured in the troposphere in the 8.5-12 bar region by the GPMS were instrumentally generated.

We suggest additional problems with the adsorption/desorption hypothesis. In the Hunten[169] model, a mole fraction of 3 ppm of non-methane hydrocarbons (mostly ethane) must be removed from ∼30 mbar (see his Fig. 3) through adsorption. For a 30-mbar atmospheric density of $1.6 \times 10^{18}$ cm$^{-3}$, that means $4.8 \times 10^{12}$ ethane molecules cm$^{-3}$ must be removed, for a column density of $9.6 \times 10^{18}$ ethane molecules cm$^{-2}$ (or really only ∼70% of this value since some ethane gas remains at 30 mbar), compared with an estimated stratospheric haze column density of $(3-8) \times 10^8$ particles cm$^{-2}$ (see Fig. 12 of West *et al.*[172]). These haze particles must be fluffy indeed to accommodate ∼ $10^{10}$ adsorbed ethane molecules per particle. The situation is even worse at 1 bar, where an ethane mole fraction of ∼$4 \times 10^{-7}$ must be lost, corresponding to an ethane column density of ∼$4.4 \times 10^{19}$ cm$^{-2}$ that must removed from the 1-bar region. The total mass that must be adsorbed per particle is likely greater than the mass of the particle itself, and layer upon layer of ethane molecules must adsorb on top of each other, which does not typically happen in adsorption processes. In fact, Curtis *et al.*[173] show that significant ethane adsorption (*i.e.*, a monolayer or greater) does not occur on tholin particles unless the ethane concentration is at or above the ethane saturation vapor density, which never occurs in the Jovian atmosphere. Although the stratospheric hazes on Jupiter may be composed of solid $H_2O$ (from an external source), benzene, and/or butane,[50] or perhaps $P_2H_4$ and $N_2H_4$ particles that have been transported through the tropopause from below, rather than tholins, the Curtis *et al.*[173] study suggests that ethane and probably acetylene condensation under the greatly subsaturated conditions on Jupiter is not likely to be as efficient as Hunten[169] have assumed. Therefore, because of problems with the physics of the adsorption process itself, as well as to the interpretation of the GPMS data, we deem it unlikely that the Hunten aerosol adsorption/desorption mechanism[169] could operate to release a significant amount $C_2H_2$ into the Jovian troposphere.

Although it is clear that $C_2H_2$ is not abundant in the troposphere of Jupiter in a global sense, the possibility that localized regions could contain enhanced tropospheric acetylene due to lightning[24,25] or dynamical effects has not been ruled out. We therefore investigate a photochemical model for which the bottom boundary condition for $C_2H_2$ has been changed to a mole fraction of $1.5 \times 10^{-7}$.[132,133] We also change the bottom boundary conditions for $N_2$ and $CH_3NH_2$ to mole fractions of $3.52 \times 10^{-5}$ and $1.52 \times 10^{-12}$, respectively, to reflect the deep-tropospheric quenched source from our nominal thermochemical kinetics and transport model (see Fig. 4). Species like $NH_3$, $NH_2$, and $N_2H_4$ are relatively unaffected by the increased $C_2H_2$ and $CH_3NH_2$ abundances ($N_2$ is effectively inert in the Jovian upper troposphere), but $CH_2NH$, $CH_3NH_2$, HCN, $CH_3CN$, $C_2H_5NH_2$, $C_2H_3CN$, and all the nitrogen-bearing organic species observed in the Keane *et al.*[39] $C_2H_2/NH_3/H_2$ photolysis experiments are significantly enhanced when the $C_2H_2$ abundance is increased. Ethylamine, in particular, shows a particularly large increase and approaches 1 ppb in our model; however, ethylamine condensation, which is not currently included in the model, will eventually limit the column abundance. Figure 6 shows the enhancement in the HCN abundance that would result from coupled $C_2H_2$-$NH_3$ photochemistry when more $C_2H_2$ is available in the Jovian troposphere. The resulting increase in the tropospheric HCN mole fraction of two orders of magnitude is not as large as one might expect because significant loss processes for $C_2H_2$ still exist in the lower stratosphere and tropopause region that convert $C_2H_2$ to $C_2H_6$ and heavier hydrocarbons rather than to nitriles and other nitrogen-bearing organics. The $C_2H_2$ mole fraction remains low in the $NH_3$ photolysis region, with a minimum mole fraction of only 1.4 ppb at 124 mbar. The HCN mole fraction therefore still remains below the observational upper limits, even with the increased $C_2H_2$ lower boundary condition. The only way to increase the HCN abundance from coupled $NH_3$-$C_2H_2$ coupled chemistry would be to invoke rapid dynamical mixing near the tropopause, such that more $C_2H_2$ and $NH_3$ are available in the ammonia photolysis region.

## 5.3. Thunderstorm source for HCN

Bar-Nun and Podolak[24] and Podolak and Bar-Nun[25] advocate lightning, and the resulting shock waves produced from lightning, as a source of HCN (and $C_2H_2$ and CO) for the Jovian troposphere. A discussion of the relative effectiveness of such a source is beyond the scope of this paper. The Bar-Nun and Podolak[24] and Podolak and Bar-Nun[25] model calculations depend on several unknown parameters such as production efficiencies per energy released in the lightning event, the location of lightning within the cloud (which must be in the top portion of the $H_2O$ cloud to have HCN and $C_2H_2$ as significant products), the assumed attenuation of the observed visible light within the cloud, the optical efficiency of the lightning, and the fate of the products once they are generated. The fact that their derived HCN mole fraction is greater than the observational upper limits[13–15] suggests that some of the terms in their calculation may need reevaluation. Note that although lightning production of $C_2H_2$ could lead to enhanced photochemical production of HCN, the resulting HCN mole fraction would not necessarily be observable (see Fig. 6).

Several observational tests should be available to evaluate the likelihood of lightning contributing to the global production of disequilibrium species on Jupiter. If production is as effective as the models suggest,[24,25] disequilibrium species should be greatly enhanced within active thunderstorms, and the dispersion to other latitudes would not be instantaneous. Therefore, investigators could look for local enhancements in disequilibrium tropospheric species (*e.g.*, $C_2H_2$ from thermal infrared or ultraviolet observations, CO and HCN from infrared, millimeter, or sub-millimeter observations) at latitudes where active lightning storms are known to be prevalent.[174] Such observational tests might be more readily available for Saturn, from Cassini data already obtained or from the extended mission, but should also be possible for Jupiter from ground-based observations and from the Juno mission. We encourage studies of the spatial distribution of tropospheric disequilibrium constituents on Jupiter and Saturn to help evaluate the lightning source.

## 6. Conclusions

We have developed two theoretical models to investigate the production and

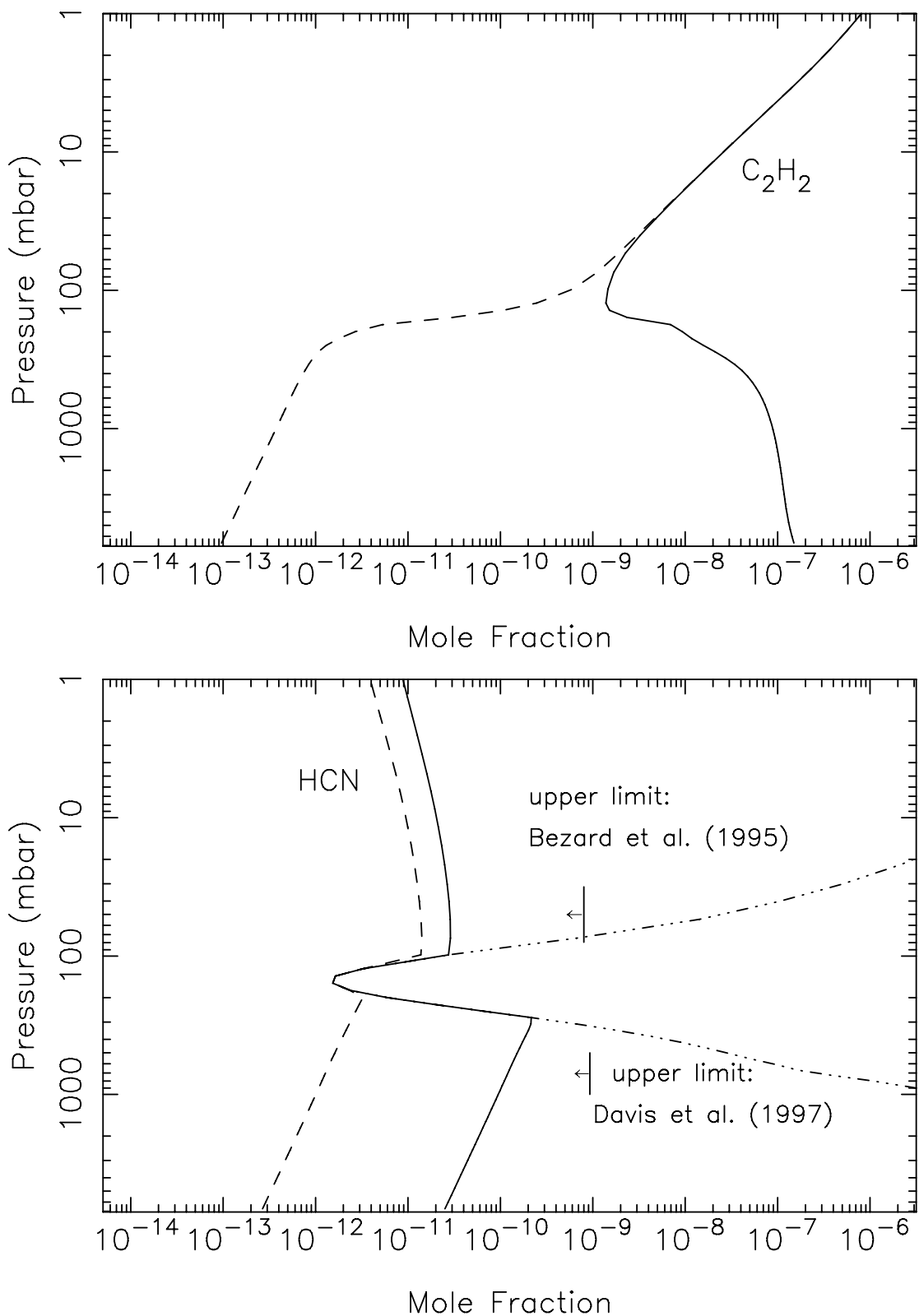

FIGURE 6. Mole-fraction profiles for acetylene (top) and hydrogen cyanide (bottom) from our Jovian photochemical model. The dashed line represents our nominal model, as discussed in Section 3. The solid line represents a model in which the bottom boundary conditions for $C_2H_2$, $N_2$, and $CH_3NH_2$ have been increased to mole fractions of $1.5 \times 10^{-7}$, $3.52 \times 10^{-5}$, and $1.52 \times 10^{-12}$.

loss of HCN and other nitrogen-bearing organics in the atmosphere of Jupiter. The first model covers the upper troposphere and stratosphere, for which we use the Caltech/JPL KINETICS code[43] to track coupled $NH_3$-$C_2H_2$ and $NH_3$-$CH_4$ photochemistry. The second model covers the deep troposphere, for which we again use the KINETICS code, but this time to track the thermochemical kinetics and transport of nitrogen species in the hot, high-pressure troposphere. We use simulations of the laboratory photolysis experiments,[38,39] along with theoretical calculations from the combustion-chemistry literature,[60] to help constrain uncertain rate coefficients in our reaction mechanism. We find that the photochemical production of HCN and other organo-nitrogen compounds is greatly inhibited in Jupiter's atmosphere. As was first discussed by Strobel,[100] our models suggest that ammonia condensation in the upper troposphere, combined with efficient photolysis and the resulting generation of the condensible $N_2H_4$ photoproduct, limit the availability of $NH_3$ in the Jovian stratosphere. Ammonia then does not diffuse into the upper atmosphere to the methane photolysis region to participate in coupled $NH_3$-$CH_4$ photochemistry. Contrary to several suggestions in the literature,[35−39] we find that coupled $NH_3$-$C_2H_2$ photochemistry is inefficient in Jupiter's troposphere when realistic $C_2H_2$ mixing-ratio profiles are considered. As is apparent from both photochemical models and observations, acetylene has a large mixing-ratio gradient in the stratosphere, leading to low abundances of $C_2H_2$ in the tropopause region. The main factor inhibiting the photochemical production of HCN and other nitrogen-bearing organics on Jupiter is the low acetylene abundance in the region where $NH_3$ is being photolyzed. Our predicted HCN mole fraction from our photochemical model is well below the upper limits derived from infrared and sub-millimeter observations.[13−15]

Consistent with other investigations that were based solely on time-constant arguments,[10,20,21,145] our thermochemical kinetics and transport models suggest that transport-induced quenching of equilibrium abundances in Jupiter's deep troposphere leads to large predicted mole fractions of $N_2$ and small predicted mole fractions of hydrogen cyanide in Jupiter's upper troposphere. However, our mechanisms for HCN and $N_2$ destruction differ considerably from those suggested by previous investigators. Our models confirm the results of Smith,[46] who demonstrates that the effective length scale for atmospheric mixing has been overestimated in the time-scale arguments of the above investigators. As a result, the suggested rate-limiting step for HCN destruction originally suggested by Fegley and Prinn[147] and used by later investigators is too slow to be consistent with the observed HCN upper limits — HCN would quench too deep in the atmosphere, where the equilibrium HCN mole fraction is large, if the suggested HCN + $H_2 \rightarrow CH_2 + NH$ reaction is the rate-limiting step. We instead find that HCN in our thermochemical kinetics and transport model is destroyed through a series of reactions that begin with H-atom addition to HCN, followed by reactions with $H_2$ and H to eventually form a single-bonded C-N species ($CH_3NH_2$) that can then thermally decompose to break the C-N bond (see scheme (2) above). The rate-limiting step for HCN reduction in our mechanism is the reaction $H_2 + H_2CN \rightarrow CH_2NH + H$, where we have calculated the rate coefficient of this reaction from that of the reverse reaction.[60] This scheme, and others like it, are much more efficient than the proposed HCN + $H_2$ destruction reaction,[147] and our predicted quenched disequilibrium HCN mole fraction is therefore comfortably below the HCN observational upper limits.[13−15]

Reduction of $N_2$ in our model follows a similar scheme (see scheme (3) above) with the reaction $N_2H_2 + H + M \rightarrow N_2H_3 + M$ being the rate-limiting step. Molecular nitrogen is likely to be very abundant ($\sim$30 ppm) on Jupiter from this quenched deep-tropospheric source (see also Prinn and Olaguer[147] and subsequent modelers), and $N_2$ should survive to be transported up into the upper atmosphere, where its mole fraction will eventually be reduced due to molecular diffusion. The interaction of galactic cosmic rays with $N_2$ might initiate interesting $N_2$ chemistry on Jupiter (see the equivalent Neptune study[155]), although the stratospheric HCN upper limit of 0.8 ppb[13] may place constraints on the effectiveness of this process. Molecular nitrogen may also participate in Jovian ionospheric chemistry, as on Titan,[156,175−181] with interesting consequences for the composition, structure, and time-variability of the lower ionosphere, as well as for the production of polycyclic aromatic hydrocarbons and other neutral species. The chemistry of $N_2$ in the auroral regions might be particularly interesting. The suggested large $N_2$ mole fraction on Jupiter needs observational confirmation, however.

One way to investigate the hypothesis of $\sim$30 ppm $N_2$ on Jupiter would be to take a more detailed look at the Galileo Probe Mass Spectrometer data. The $N_2$ abundance has never been definitively determined from the GPMS data because of concerns over an internal source of $CO_2$ from the instrument itself,[110] which could contaminate the signal at 28 dalton/$e^-$ due to the $CO^+$ daughter ion of $CO_2$.[182] Wong[171] calculated a combined CO + $N_2$ mixing ratio relative to $H_2$ of $2.3 \times 10^{-7}$ in the 8.5-12 bar pressure region of the probe entry site. An $N_2$ mixing ratio greater than this value can only be supported if the contribution from the internal instrumental $CO_2$ source was overestimated in the Wong[171] study;[182] this uncertainty highlights the need for further calibration studies to better characterize the instrument-generated $CO_2$ signal (if any) in the GPMS data.

The observational upper limits for tropospheric HCN[13-15] and the lack of $C_2H_2$ absorption wings in thermal-infrared data[133,161] can be used help constrain the effectiveness of other disequilibrium production processes, such as that of lightning-induced chemistry.[24,25] Upper limits to the tropospheric $C_2H_2$ mole fraction from thermal-infrared observations are seldom provided in the literature, although such information would aid our understanding of the role of lightning on Jupiter and the other giant planets. Spatially resolved observations that compare the tropospheric composition of latitude regions known to have active thunderstorms with the composition of quiescent regions would also provide important information on this topic.

The mechanisms proposed in our investigation are still speculative due to the limited experimental data available for individual reactions of interest in the models. A better understanding of the pathways for reduction of HCN and $N_2$ is of importance for studies of the composition and chemical behavior of extrasolar giant planets and brown dwarfs, as well as for giant planets within our own solar system, and further investigation into these processes is warranted. Our thermochemical kinetics and transport model in particular can be applied to studies of the "hot Jupiters" that are being discovered at an astonishing rate around other stars. The photochemical model may also be of use to extrasolar-giant-planet studies. Although we find that coupled carbon-nitrogen photochemistry is not important on Jupiter, that result is largely due to the removal of $NH_3$ from the upper atmosphere due to condensation, and we may anticipate alternative scenarios for warmer giant planets. If an extrasolar giant planet were located closer to its parent star (such that $NH_3$ does not condense) but not so close that it is being intensely bombarded by ultraviolet radiation (such that strongly bonded species are not the only surviving molecules), then coupled $NH_3$-$C_2H_2$ and $NH_3$-$CH_4$ photochemistry could be very important indeed (see the theoretical planetary classes of Sudarsky *et al.*[183]). Coupled $NH_3$-$C_2H_2$ photochemistry might also be important in cometary comae or any other astronomical environment where ammonia and acetylene are brought together in the presence of ultraviolet radiation. We therefore encourage further investigation into the thermodynamic and kinetic properties of the organo-nitrogen compounds observed in the photolysis experiments of Ferris and Ishikawa,[36,37] Keane,[38] and Keane *et al.*[39]

# Acknowledgments

The Caltech/JPL KINETICS code was developed jointly by Yuk L. Yung and Mark Allen, with assistance from many people over the years, and we thank them for letting us continue to use this powerful and flexible code. Veronica LaMothe helped with initial forays into Jovian tropospheric photochemistry. We thank Michael H. Wong for useful discussions, and we gratefully acknowledge financial support from the NASA Planetary Atmospheres Program (NNX08AF05G for the photochemistry portion, NNX09AB55G for the thermochemistry portion).